\documentclass{article}

\usepackage{arxiv}

\usepackage[utf8]{inputenc} 
\usepackage[T1]{fontenc}    
\usepackage{hyperref}       
\usepackage{url}            
\usepackage{booktabs}       
\usepackage{amsfonts}       
\usepackage{nicefrac}       
\usepackage{microtype}      
\usepackage{lipsum}		
\usepackage{graphicx}
\usepackage[numbers]{natbib}
\usepackage{doi}
\usepackage{caption}
\usepackage{subcaption}

\title{HaDR: Applying  Domain Randomization for Generating Synthetic Multimodal Dataset for Hand Instance Segmentation in Cluttered Industrial Environments}

\author{ \href{https://orcid.org/0000-0002-8984-153X}{\includegraphics[scale=0.06]{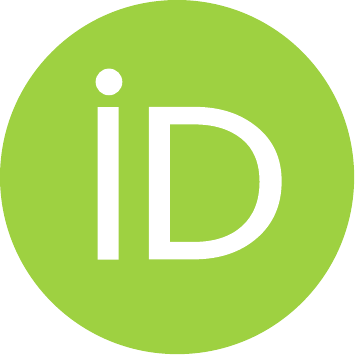}\hspace{1mm}Stefan Grushko}
\thanks{
Dataset and pretrained models are available in \href{https://doi.org/10.34740/KAGGLE/DS/2970535}{https://doi.org/10.34740/KAGGLE/DS/2970535}.\newline
\indent Code is available in \href{https://github.com/anion0278/HaDR}{https://github.com/anion0278/HaDR}.
} \\
	Department of Robotics,\\ Faculty of Mechanical Engineering,\\ VSB - Technical University of Ostrava\\
	\texttt{stefan.grushko@vsb.cz} \\
	\And
	\href{https://orcid.org/0000-0001-6942-4280}{\includegraphics[scale=0.06]{orcid.pdf}\hspace{1mm}Aleš Vysocký} \\
	Department of Robotics,\\ Faculty of Mechanical Engineering,\\ VSB - Technical University of Ostrava\\
	\texttt{ales.vysocky@vsb.cz} \\
 \And
	\href{https://orcid.org/0000-0002-4946-8638}{\includegraphics[scale=0.06]{orcid.pdf}\hspace{1mm}Jakub Chlebek} \\
	Department of Robotics,\\ Faculty of Mechanical Engineering,\\ VSB - Technical University of Ostrava\\
	\texttt{jakub.chlebek@vsb.cz} \\
 \And
	\href{https://orcid.org/0000-0002-5881-2779}{\includegraphics[scale=0.06]{orcid.pdf}\hspace{1mm}Petr Prokop} \\
	Department of Computer Science,\\ Faculty of Electrical Engineering\\ and Computer Science,\\ VSB - Technical University of Ostrava \\
	\texttt{petr.prokop@vsb.cz} \\
}

\renewcommand{\shorttitle}{HaDR: Domain Randomization for Hand Instance Segmentation}

\hypersetup{
pdftitle={HaDR: Domain Randomization for Hand Instance Segmentation},
pdfauthor={S. Grushko, A. Vysocky, J. Chlebek, P. Prokop},
pdfkeywords={instance segmentation, hand detection, synthetic dataset, domain randomization, deep learning, multimodal},
}

\begin{document}
\maketitle
\begin{abstract}
	This study uses domain randomization to generate a synthetic RGB-D dataset for training multimodal instance segmentation models, aiming to achieve colour-agnostic hand localization in cluttered industrial environments. Domain randomization is a simple technique for addressing the "reality gap" by randomly rendering unrealistic features in a simulation scene to force the neural network to learn essential domain features. We provide a new synthetic dataset for various hand detection applications in industrial environments, as well as ready-to-use pretrained instance segmentation models. To achieve robust results in a complex unstructured environment, we use multimodal input that includes both colour and depth information, which we hypothesize helps to improve the accuracy of the model prediction. In order to test this assumption, we analyze the influence of each modality and their synergy. The evaluated models were trained solely on our synthetic dataset; yet we show that our approach enables the models to outperform corresponding models trained on existing state-of-the-art datasets in terms of Average Precision and Probability-based Detection Quality.
\end{abstract}


\section{Introduction}
When it comes to manufacturing process flexibility, robot programming can be a barrier to customization and reconfiguration. In practice, programming is usually done through interaction with the robot control panel or through workspace simulation. Programming an industrial robot nowadays considers human-centred interaction process design \cite{PAN2023102492} and often employs human-human interaction as a model of inspiration for modalities of human-robot interaction such as gestures, speech, and gaze \cite{LI2023102510, mti7030025}. Interfaces based on hand gestures may be considered more natural and easier to use for human users, who can use them in many tasks, not limited to high-level robot action control (but also augmented reality, virtual reality, and general human-computer interaction), because gestures are a common interaction that humans naturally use in social communication. Typical approaches to gesture recognition and hand tracking utilize data collection gloves, and various markers \cite{KIM2020101819, AMORIM2021102035}, however, most approaches focus on vision-based systems since it does not require any additional equipment for the user to work with. 

Machine learning, and in particular deep learning (DL) approaches, have made it possible to achieve high accuracy, robustness, and have generally made recognition models more accessible, provided that an appropriate amount of data is used to train the model. In computer vision, instance segmentation is the typical task of performing pixel-level segmentation of individual instances (as the opposite to semantic segmentation). It is a more challenging and demanding task than object detection because it requires both instance-level and pixel-level predictions. However, there have been notable advancements in recent times towards improving the performance and accuracy of instance segmentation. Nevertheless, training these systems from scratch is still a challenge since these methods depend on the availability of large high-quality datasets, whose annotation requires expensive pixel-level segmentation due to the amount of time required for a human to manually label a single sample.

One possible solution to address this limitation is to use graphical simulation platforms to generate large sets of automatically labelled samples without human intervention. Yet, in order to successfully use these simulated environments in an autonomous generation process, they must first be manually prepared, which is a tedious and expensive task because it requires careful modelling of specific environments with high attention to photorealistic details. Consequently, the cost required to generate photorealistic environments undermines the main advantage of synthetic data, namely the generation of arbitrarily large amounts of labelled data. Domain Randomization (DR) is a methodology aimed at lowering the expenses associated with generating large quantities of precisely labelled data. This approach involves deliberately disregarding photorealism and instead introducing non-photorealistic variations to the environment through random perturbations, such as adding random textures, light sources, and distracting objects. The goal is to force the network to generalize to essential domain features.

The motivation for this work was to mitigate the enduring issue of publicly available DL solutions, which often suffer from degradation since they are mostly trained on real-world data, biasing the network towards relying on texture and hand skin colour (such as MediaPipe \cite{zhang_mediapipe_2020}). Our goal was to provide a fully synthetic dataset that would allow training the DL model in a colour-agnostic manner without using real camera data. Our intended field of application is an unstructured industrial environment since our goal is to further incorporate the developed system as part of a human-robot interaction interface. As the colour of work gloves (which are often part of personal protective equipment \cite{LUCCI2022102384}) and the background can often be difficult to distinguish from RGB data alone, we explore multimodal input that includes colour and depth information about the scene and the influence of the components of this input, as opposed to our previous work where only depth data was used to segment the image \cite{vysocky_generating_2022}. We focus on the instance segmentation task because it will further allow to aggregate and filter all points related to the user's hand and then combine these point clouds from multiple cameras to evaluate the shape and position of the hand more accurately (which would not be possible in the case of hand landmark recognition). To the best of the authors' knowledge, this is the first attempt to create a colour-agnostic RGB-D synthetic dataset that can mitigate the issue of over-reliance on human skin features. 

Instance segmentation models trained solely on our synthetic dataset outperform the corresponding models trained on state-of-the-art real and synthetic datasets by achieving up to 52.5 AP (Average Precision, COCO’s challenge standard metric IoU at 0.5:0.95 \cite{lin_microsoft_2014}, which is assumed throughout the text unless otherwise stated). Our models also outperform the MediaPipe hands solution in the bounding box detection task in both AP and Probability-based Detection Quality (PDQ) \cite{hall_probabilistic_2020} metrics on a challenging dataset representing an unstructured industrial environment, demonstrating independence from the colour of the work gloves worn on the hands. 

This paper is organized as follows: in Section 2, we review related work on dataset collection techniques as well as existing datasets adapted for the hand detection task. In Section 3, we explain the methodology, which consists of the setup of the simulated scene, the process of collecting a multimodal dataset, and the training of the instance segmentation models. Section 4 discusses the results obtained using DR and the importance of different input components. It also evaluates the existing datasets and MediaPipe Hands solution in order to compare them with our DR dataset. Finally, Section 5 analyzes the results and limitations, and draws conclusions about the presented solution while also providing future research directions.

\newpage
\section{Related Work}
Recent advances in DL have boosted research and produced many breakthroughs in computer vision and machine learning in general. However, supervised training of deep neural networks is heavily dependent on the availability of a sufficiently large, domain-specific and properly-labelled dataset, which is time-consuming and generally complicated to be prepared manually. Several simplifying approaches have been proposed to streamline the process of collecting large datasets. In the case of the hand recognition domain, the hand detection task itself can be simplified using various (visual, magnetic) tags \cite{hillebrand_inverse_2006, wetzler_rule_2015, KIM2020101819, AMORIM2021102035}, sensor-equipped gloves (data gloves)\cite{baldi_gesto_2017, grushko_intuitive_2021}, specialized sensors \cite{vysocky_analysis_2020}. Another alternative is represented by synthetic approaches in which each sample is constructed by combining a known ground truth with a random background \cite{MAZHAR201934}, or by generating a complete sample using a Generative Adversarial Network (GAN) \cite{yurtsever_photorealism_2022, mueller_ganerated_2018} or simulation environment \cite{romero_embodied_2017, zimmermann_learning_2017}. 
\subsection{Synthetic dataset generation}
Synthetic datasets have gained in popularity in recent years since they offer a way to generate arbitrarily large accurately labelled datasets. The commonly used option is to incorporate renderers or full-fledged simulators into the dataset acquisition process. These visualization tools may generate the whole image or only the objects of interest and utilize a real photo as a background. Dwibedi et al. \cite{dwibedi_cut_2017}, Georgakis et al. \cite{georgakis_synthesizing_2017} proposed an alternative to complete image synthesis within a simulation but instead embedding real object images on a set of randomly selected real background images. 

Synthetic data can have a "reality gap" \cite{tobin_domain_2017} with the real world due to limitations in fully replicating real-world data, such as textures, lighting, and complex domain specifics \cite{ganin_unsupervised_2015, liebelt_multi-view_2010}. Realistic simulations are only able to cover a user-defined scale of conditions, such as specific lighting conditions or a limited subset of object positions and interactions. As a result, the generated environments may only represent a portion of the many conditions that can occur in reality. There are two common approaches to overcoming this disparity: either reducing the "reality gap" by trying to increase the similarity between the simulated and real environments \cite{planche_physics-based_2021} or focusing on features that are specific and important for generalization. 

Achieving photorealism using high-fidelity rendering engines comes at the cost of computing resources and rendering time, as well as costs associated with manual synthetic scene layout, environment design, shader and effect adjustments, and implementation of domain rules within the simulated model. Several research groups \cite{yurtsever_photorealism_2022, mueller_ganerated_2018, oprea_h-gan_2021} have attempted to use GANs to improve the photorealism of synthetic images as an alternative to high-fidelity rendering. Mueller et al. \cite{mueller_ganerated_2018} utilized the image-to-image translation model CycleGAN \cite{zhu_unpaired_2020} to transform synthetic hand images generated in a simulation into more photorealistic representations while ensuring geometric consistency. GAN-based approaches however imply additional effort required to train GAN.

As an alternative, the latter approach embraces the reality gap and instead applies additional disturbances to features that the machine learning model should generalize rather than trying to perfectly mimic reality. Sadeghi and Levine \cite{sadeghi_cad2rl_2017} extended the known domain adaptation \cite{tzeng_adapting_2017} approach and proposed a fully simulated vision-based policy training for quadcopter indoor navigation and collision avoidance tasks where a simulation environment was used without relying on visual fidelity. The term domain randomization was introduced in the works of Tobin et al. \cite{tobin_domain_2017}, where the researchers attempted to mitigate the issue of the "reality gap" by generating unrealistic synthetic RGB training data with sufficient variability in the domain features. Variable parameters included random number, position and shapes of distractor objects, the texture of all objects and background, the position, orientation, and field of view of the camera and lights in the scene, and the type and amount of random noise added to the images. This results in a dataset with a wide distribution of features, as opposed to one that might be observed when manually collecting real-world data, which helps to increase robustness to high environmental variability. Hinterstoisser et al. \cite{hinterstoisser_pretrained_2017} combined DR with synthetic images generated by combining real backgrounds with overlayed rendered objects of interest, while these objects were randomized with use of random noise, illumination, and blurring. Tremblay et al. \cite{tremblay_training_2018} used the DR approach to create a large dataset for training the object detection model, which reduced the time required to prepare the dataset and also provided better results than using photorealistic datasets or real data alone. Similar approaches to generating datasets were applied by Dehban et al.\cite{dehban_impact_2019}, Khirodkar et al. \cite{khirodkar_domain_2018}, Horvath et al. \cite{horvath_object_2022} for the tasks of object detection in diverse domains. 

In many cases \cite{tremblay_training_2018, mueller_real-time_2019}, synthetic datasets are complemented and fine-tuned with real-world images to achieve better results, however in contrast, we use synthetic data to train networks that segment instances of real-world objects based only on the synthetic dataset while achieving stable results.

\subsection{Hand datasets}
Here we focus on the domain of hand detection by covering datasets for semantic and instance segmentation, key point detection, and pose estimation. 

For hand detection and segmentation, typical datasets consist of manually annotated images obtained with a real RGB camera. The EgoHands dataset \cite{bambach_lending_2015} includes 4.8K images of two people's interactions, where  pixel-level segmentation masks for each hand were manually annotated. Nuzzi et al. presented an RGB-D HANDS dataset \cite{nuzzi_hands_2021, NUZZI2021102085} comprising of samples with single hand gestures (only gesture classification was provided, and no segmentation masks were created).

Automatically annotated datasets may significantly simplify hand detection and segmentation using markers or coloured gloves. An example of a such dataset was presented by A. Bojja et al. \cite{bojja_handseg_2018} with the HandSeg dataset containing 150K samples with random hand gestures in front of a depth camera (the ground truth annotations were created automatically using colour gloves and HSV thresholding). Real camera datasets are advantageous as they capture realistic domain features. However, it is worth noting that they typically represent only a subset of environmental conditions and, as a result, they cannot cover the full range of scenarios, such as changes in obstacle positions, illumination, reflections, and shadows. This bias towards the captured conditions may affect the resulting model.

Fully synthetic datasets generated using renderers or simulators are advantageous due to the ease and speed with which they can acquire perfectly aligned multimodal samples. However, one of their limitations is their inability to  represent domain-specific features accurately and realistically. Mueller et al. \cite{mueller_real-time_2019} demonstrated pose and shape reconstruction of interacting hands using a model trained on a synthetic dataset (Dense Hands). The dataset consisted of depth image samples augmented with RGB-encoded segmentation masks representing the vertices of a MANO hand model \cite{romero_embodied_2017}. To overcome the limitations of the generated dataset (absence of background obstacles and augmentations), the model was also trained on real camera data to improve its generalization.
Combining RGB and depth modalities provides more information to the segmentation and increases the possible variability of the generated dataset. Zimmermann and Brox presented RGB-D Rendered Hand Pose dataset (RHD) as part of their work dedicated to the estimation of 3D hand pose from camera image \cite{zimmermann_learning_2017}. They employed 3D character models matched with a highly parameterizable hand 3D model  MANO \cite{romero_embodied_2017}. Each dataset frame contained a randomly posed human character model (using realistic animation), while the camera position was randomly selected from a vicinity surrounding the hand of the character. The generated scenes included randomized backgrounds, global lighting, specular reflections, and directional light sources. However, the possible range of hand positions was limited, so the hand only appeared in the centre of the image and never at the edges. ObMan dataset \cite{hasson_learning_2019} is a fully synthetic dataset consisting of RGB-D images created using 3D models of human characters holding various objects. The camera was randomly positioned to capture various poses, backgrounds, textures, and lighting conditions. The dataset includes 150k fully annotated images (hand key points,  segmentation masks for objects and hands).

Our strategy differs from earlier works by utilizing Domain Randomization instead of relying on photorealistic synthetic images. Additionally, we aim to address limitations present in publicly available datasets, which include:
\begin{itemize} 
\item Being limited to either RGB or Depth information \cite{bambach_lending_2015, mueller_real-time_2019};
\item Assuming that the hand occupies the majority of the image area \cite{qian_realtime_2014, tompson_real-time_2014};
\item Only considering samples with a single instance \cite{qian_realtime_2014, tompson_real-time_2014, wetzler_rule_2015};
\item Biasing instance locations toward the center of image \cite{zimmermann_learning_2017, bojja_handseg_2018, hasson_learning_2019};
\item Assuming that the hand is the closest object to the camera \cite{zimmermann_learning_2017, bojja_handseg_2018, hasson_learning_2019};
\item Lacking distractor objects and obstructions around the hand \cite{qian_realtime_2014, tompson_real-time_2014, wetzler_rule_2015, mueller_real-time_2019}.
\end{itemize}

Our decision to create a customized dataset was also driven by the fact that the currently available datasets assume a camera placement that is different from the one used in our specific industrial case. This mismatch, along with other factors, served as a motivation to generate our own dataset tailored to our specific requirements.

\subsection{Instance segmentation}
As the scope of this paper does not encompass a comprehensive examination of the instance segmentation task, we direct readers to surveys such as the work of Gu et al. \cite{gu_review_2022}. 

Instance segmentation implies the correct detection pertaining all objects in the image and at the same time semantically segmenting each instance at the pixel level. Although the objects in the image are drawn from a predefined set of semantic categories, the number of instances of these objects can vary arbitrarily. Recent approaches in this field can be clssified into three distinct groups.

In the top-down approach to instance segmentation \cite{huang_mask_2019, bolya_yolact_2019, liu_path_2018}, the first step involves detecting bounding boxes, followed by segmenting the instance mask within each of these bounding boxes.  
Mask R-CNN \cite{he_mask_2020} is the most widely recognized architecture for instance segmentation, as it expands upon the two-stage Faster R-CNN \cite{ren_faster_2016} object detector by incorporating a branch that segments object instances within detected bounding boxes.

The bottom-up approach to instance segmentation \cite{de_brabandere_semantic_2017, liu_sgn_2017, newell_associative_2017} produces instance masks by grouping pixels into a variable number of object instances within the image. This is achieved by assigning an embedding vector to each pixel, followed by offsetting pixels belonging to different instances and dragging nearby pixels within the same instance. Subsequently, a clustering post-processing step is required to separate the individual instances. 

Direct instance segmentation involves predicting both instance masks and their corresponding semantic categories in a single step, without requiring subsequent grouping processing. The SOLO \cite{wang_solo_2020} and SOLOv2 \cite{wang_solov2_2020} architectures exemplify this approach, as they directly produce instance masks and corresponding class probabilities in a fully convolutional manner without relying on bounding boxes or clustering. Unlike the bottom-up and top-down methods, direct instance segmentation does not depend on precise bounding box detection, nor does it require per-pixel embedding and grouping learning \cite{wang_solov2_2020}.

\section{Methods}
\begin{figure}[hb]
	\centering
		\includegraphics[width=14cm]{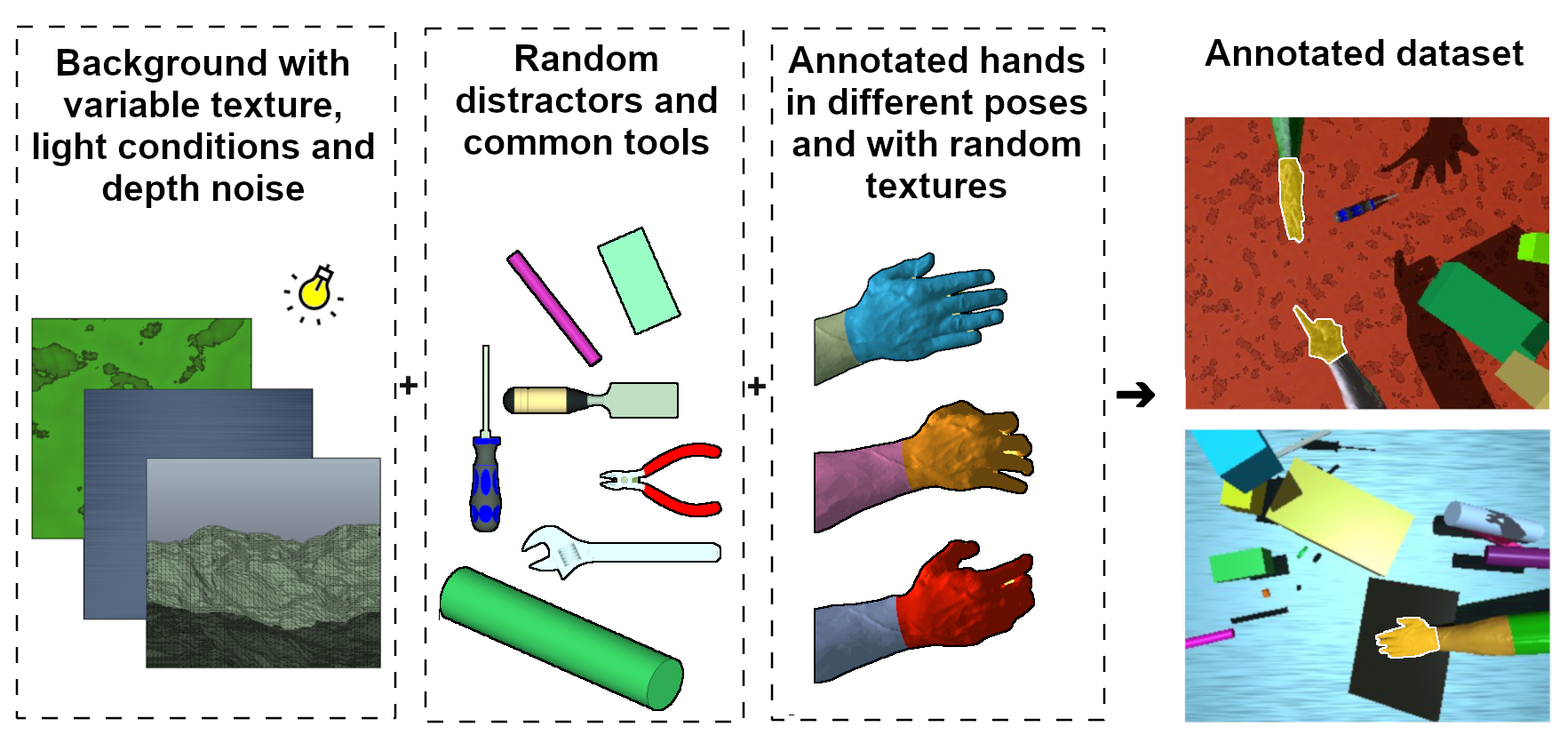}
	\caption{To enhance instance segmentation, domain randomization is implemented by overlaying annotated synthetic data onto randomized backgrounds, alongside random distractors such as geometric shapes and generic tools. The resulting scenes are then rendered with random lighting conditions, and random textures are applied to both the objects of interest and the distractor objects prior to rendering. These samples, along with automatically generated masks, are utilized for training instance segmentation models.}
	\label{fig_DR}
\end{figure}

We propose a method for training instance segmentation models using a synthetic dataset generated through simulation. The dataset is created to mimic real-world scenarios captured by a camera. To generate the synthetic dataset, we utilize domain randomization techniques, which are illustrated in Figure \ref{fig_DR}. Our dataset generator is based on the CoppeliaSim \cite{rohmer_v-rep_2013} simulation platform. The generator randomly places up to two instances of objects of interest in a 3D scene with random positions and orientations. Additionally, to improve the network's generalization to real-world scenarios, we include a variety of geometric shapes and distractors (unrelated working tools) in the scene. These shapes and distractors have random textures applied to them. Furthermore, we insert a variable number of lights at random locations in the scene. The scene is then rendered in both RGB and depth components, and the dataset generation pipeline automatically generates ground truth labels (masks) by adjusting object visibility parameters.

\subsection{Domain randomization}
Our scripted CoppeliaSim environment simplifies the generation of large-scale synthetic datasets with pixel-level accurate annotations. The conditions of the simulated scene were adjusted to match the intended use and typical camera parameters. The implemented pipeline for dataset generation produces a multimodal output, as we assume that each modality provides complementary information and aids model performance in cases where a single modality would fail (e.g., due to blending with background or insufficient depth resolution) by providing distinctive features to the networks. 
Similar to Tremblay et al. \cite{tremblay_training_2018} approach, our expectation is that generating datasets comprised of non-realistic images will force the network to differentiate the most salient features related to the shape of the objects and this, in turn, would result in better generalization to actual images.

The simulation environment employs a vision sensor to emulate a real camera, with its settings based on the field of view of a standard RGB-D camera (using values for the RealSense L515). During dataset generation, the 3D hand model is moved through a grid (20 mm increment in each axis) across the entire field of view of the camera ‐ this allows for a more uniform generation of dataset samples. The hand's orientation is semi-random, however, it follows specific policies to ensure it remains within the field of view of the vision sensor. Specifically, the fingertip point and point in the centre of palm are always within the truncated pyramid that corresponds to the camera's field of view. Additionally, the roll of the palm is limited to ±15° and the pitch to ±30°. The vision sensor captures both depth and colour images of the scene while scaling the depth pixel values in the range [0.2, 1.0] m into a single channel 8-bit range grayscale image. By our definition, random objects in the background can be closer or farther from the camera than the hand. 

The placement of these objects in the scene is also randomized with a policy in place preventing them from overlapping with the hand in the camera view. If an overlap occurs, the object is relocated until this requirement is met. Random distractors are added to the scene to reduce the sensitivity of the system to irrelevant objects that might be misidentified as fingers or hands. To avoid overfitting during model training, most textures are intentionally designed with unrealistic patterns. Separate vision sensors are responsible for capturing only the instances of the hands and outputting a single binary mask for each instance. The resulting images and masks are resized to 320x256 resolution. The samples are stored in lossless PNG format.

\begin{figure}
    \centering
\begin{subfigure}[t]{0.49\textwidth}
    \centering
    \includegraphics[height=4cm]{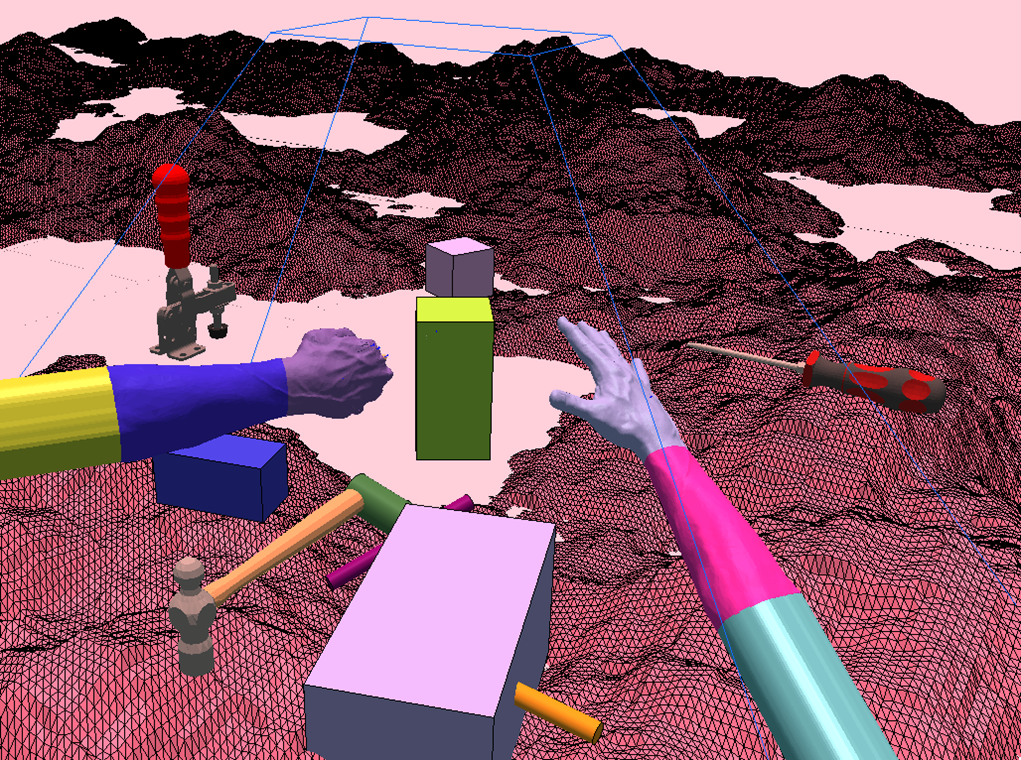}
    \caption{}
\end{subfigure}
\hfill
\begin{subfigure}[t]{0.49\textwidth}
    \centering
    \includegraphics[height=4cm]{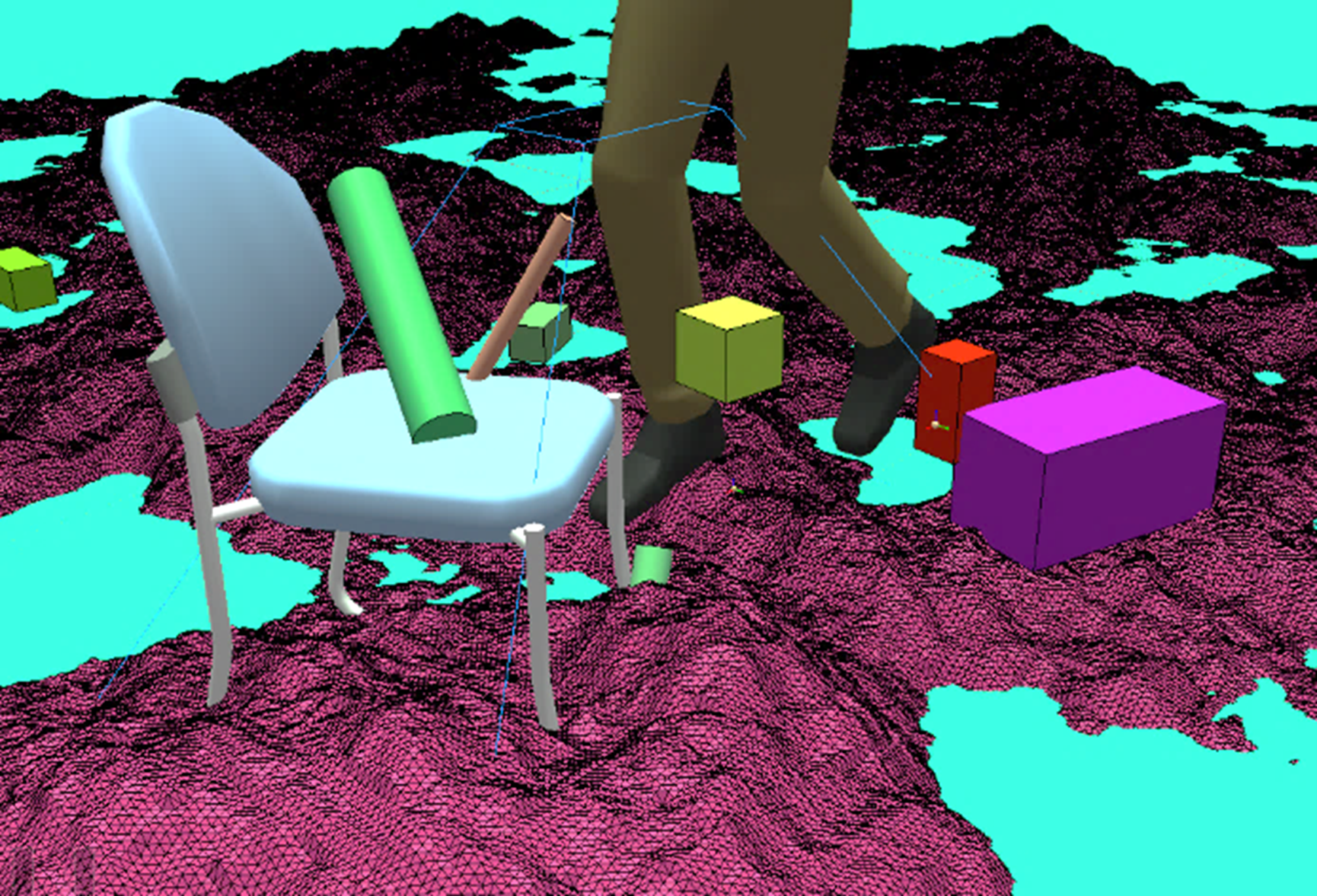}
    \caption{}
\end{subfigure}
    \caption{CoppeliaSim scene used as dataset generator: (a) scene containing two hand meshes; (b) scene containing no hand meshes and a single articulated human body model.}
    \label{fig_coppelia}
\end{figure}

\begin{figure}[hb]
	\centering
		\includegraphics[width=8cm]{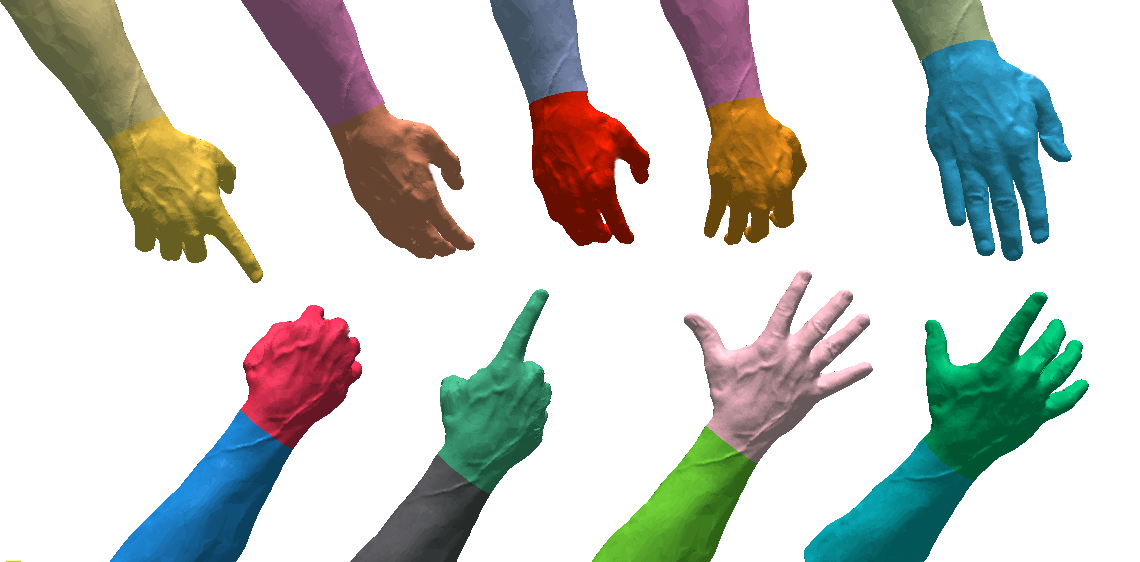}
	\caption{ Meshes with annotated hand instances used in dataset generator.}
	\label{fig_hands}
\end{figure}

The following aspects of the scene were randomly varied during the generation of the dataset at each frame:
\begin{itemize} 
\item Number, colours, textures, scales and types of distractor objects selected from a set of 3D models of general tools and geometric primitives. A special type of distractor – an articulated human body model without hands (see Figure \ref{fig_coppelia}b).
\item Hand gestures (see Figure \ref{fig_hands}).
\item Hand models’ positions and orientations.
\item Texture and surface properties (diffuse, specular and emissive properties) and number (from none to 2) of the object of interest, as well as its background. 
\item The number and positions of directional light sources, ranging from one to four, along with a planar light that provides ambient illumination.
\end{itemize}

\begin{figure} [ht]
	\centering
\begin{subfigure}[t]{0.2\textwidth}
\centering
    \includegraphics[height=8cm]{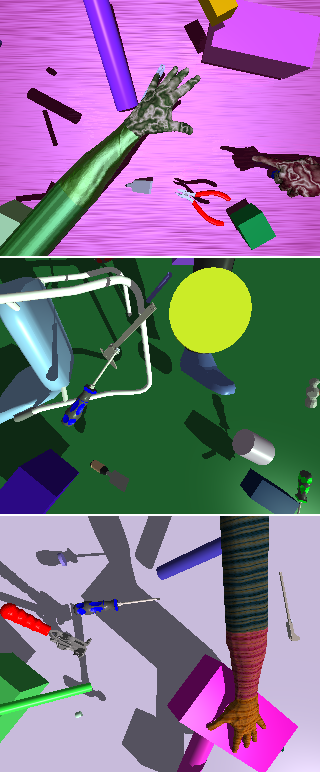}
    \caption{}
\end{subfigure}
\begin{subfigure}[t]{0.2\textwidth}
\centering
    \includegraphics[height=8cm]{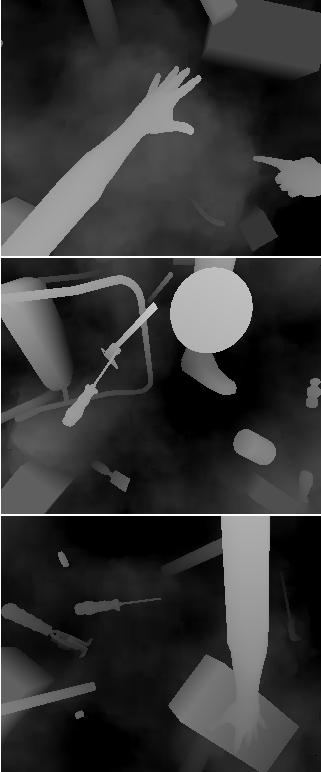}
    \caption{}
\end{subfigure}
\begin{subfigure}[t]{0.2\textwidth}
\centering
    \includegraphics[height=8cm]{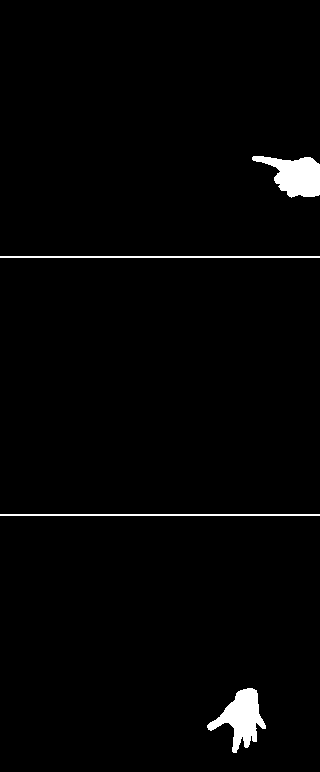}
    \caption{}
\end{subfigure}
\begin{subfigure}[t]{0.2\textwidth}
\centering
    \includegraphics[height=8cm]{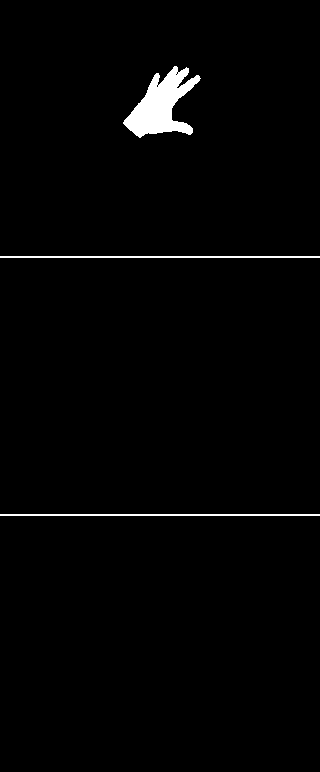}
    \caption{}
\end{subfigure}
  
	\caption{Samples from the generated dataset: (a) RGB image of the scene; (b) depth image of the scene; (c) mask of the first generated instance; (d) mask of the second instance. Generated samples include different lighting, background clutter, distractors and multiple cases where the task of hand segmentation from RGB image is complicated due to blending with the background. Samples may contain up to two instances of hands.}
	\label{fig_synth_dataset}
\end{figure}

The simulated scene contained only right hands, however, to help the models learn both sides, we used flip augmentation during training. Our dataset generation pipeline also addressed the generation of instance-free samples where only distractors are presented. The maximum number of instances per sample was limited to 2 (see Figure \ref{fig_synth_dataset}), as we assume one user in our industrial scenario. For our DR dataset, we generated a total of 117k images. Annotations in COCO \cite{lin_microsoft_2014} format were generated using PyCocoCreator \cite{wspanialy_pycococreator_2018}.

\begin{figure}[ht]
	\centering
		\includegraphics[width=10cm]{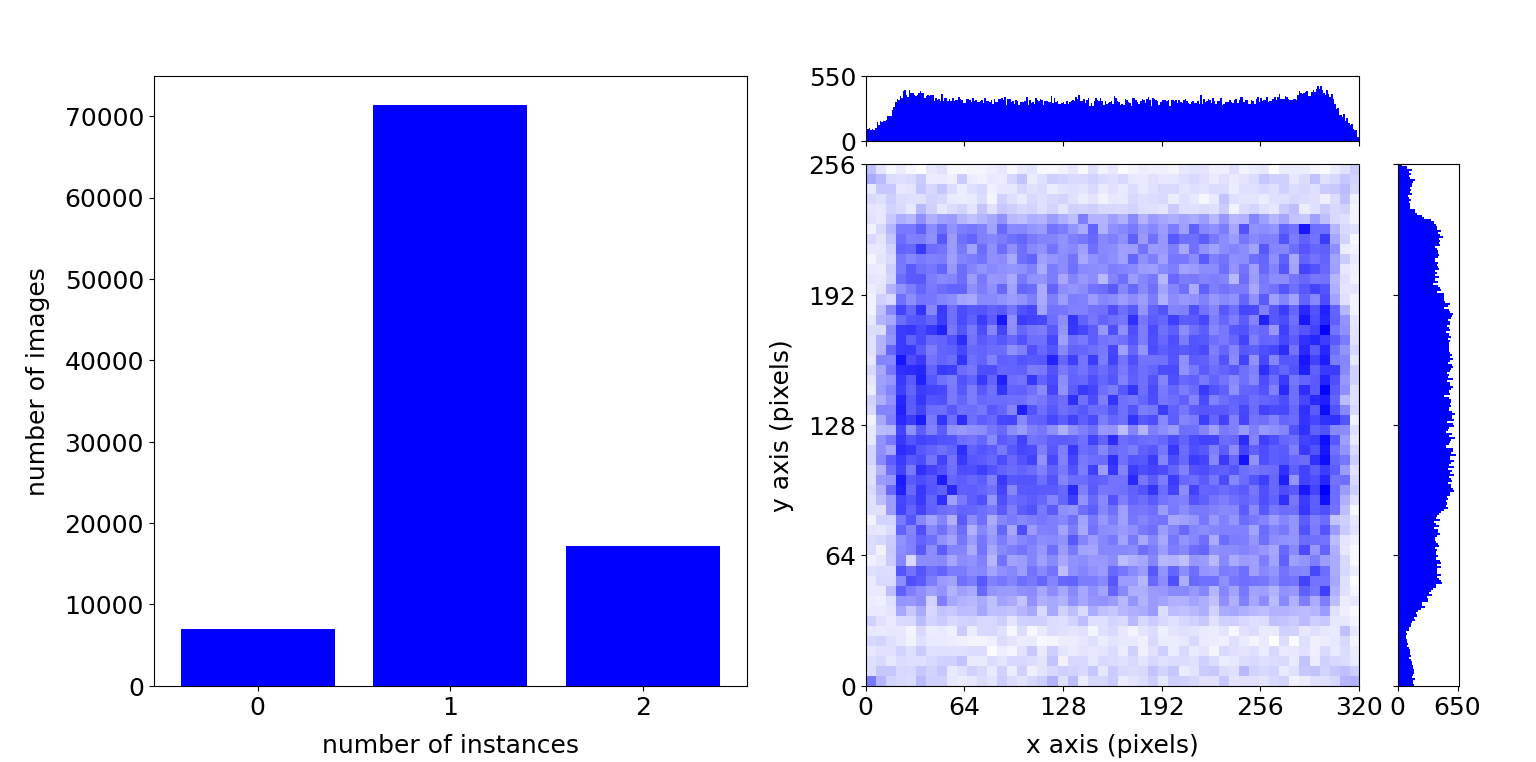}
	\caption{Statistics of our dataset: (left) shows the distributions of the number of hand instances per image and hand centroid locations (right). }
	\label{fig_statistics_our}
\end{figure}

Our dataset contains only a relatively small number of images, however since we traverse the entire available workspace in a grid during the collection process, the dataset covers the range of possibilities uniformly than other datasets (more on that in \ref{comparison_with_existing_datasets}). This is due to the fact that the collection process for other datasets typically involved random generation without using a grid-based approach; furthermore, most datasets assume that hands can only appear in a certain region of the image. The particular distribution observed in Figure \ref{fig_statistics_our} is due to the fact that we generate images in the camera field of view, which is represented by a pyramid.

\subsection{Training process}
This study adopts two state-of-the-art neural networks (each with two options of backbones: ResNet-50 and ResNet-101 as feature extractors) using their open-source implementations in MMDetection Framework \cite{chen_mmdetection_2019}. All models were trained for 20 epochs without freezing the backbone with an initial learning rate of 0.02, which was then divided by 10 at 7 and 18 epochs. The training was performed with stochastic gradient descent (SGD) on 2 GPUs with 2 images per mini-batch. We adopt standard hyper-parameters with an initial learning rate of 0.02, a learning rate decay factor of 0.1, and a weight decay of 0.001. 

To improve sensitivity for both left and right hands, we employed horizontal and vertical image flipping augmentations during training. This was necessary since the datasets were solely generated using a model of the right hand. All networks were initialized models with COCO-pretrained (MS COCO train2017 \cite{lin_microsoft_2014}) 3-channel models (ResNet backbones initially pretrained on ImageNet) to reduce the required training time and dataset size. The dataset is split into training and validation sets with an 80/20 ratio. The training was performed on a workstation with 32 GB of RAM, Intel Core i7-9700F CPU and equipped with two NVIDIA GeForce RTX 3080 Ti (12GB) graphic cards. 

\subsection{Evaluation}
\begin{figure}[hb]
	\centering
\begin{subfigure}[t]{0.2\textwidth}
    \centering
    \includegraphics[height=7.5cm]{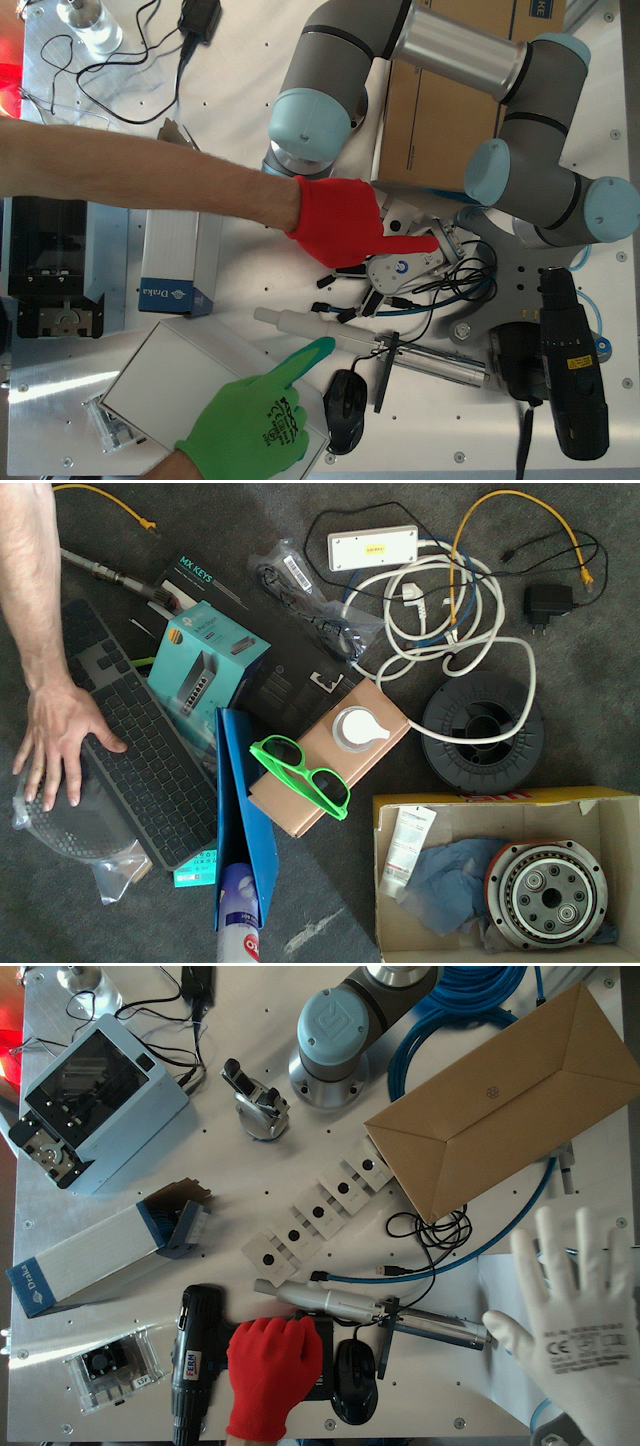}
    \caption{}
\end{subfigure}
\begin{subfigure}[t]{0.2\textwidth}
    \centering
    \includegraphics[height=7.5cm]{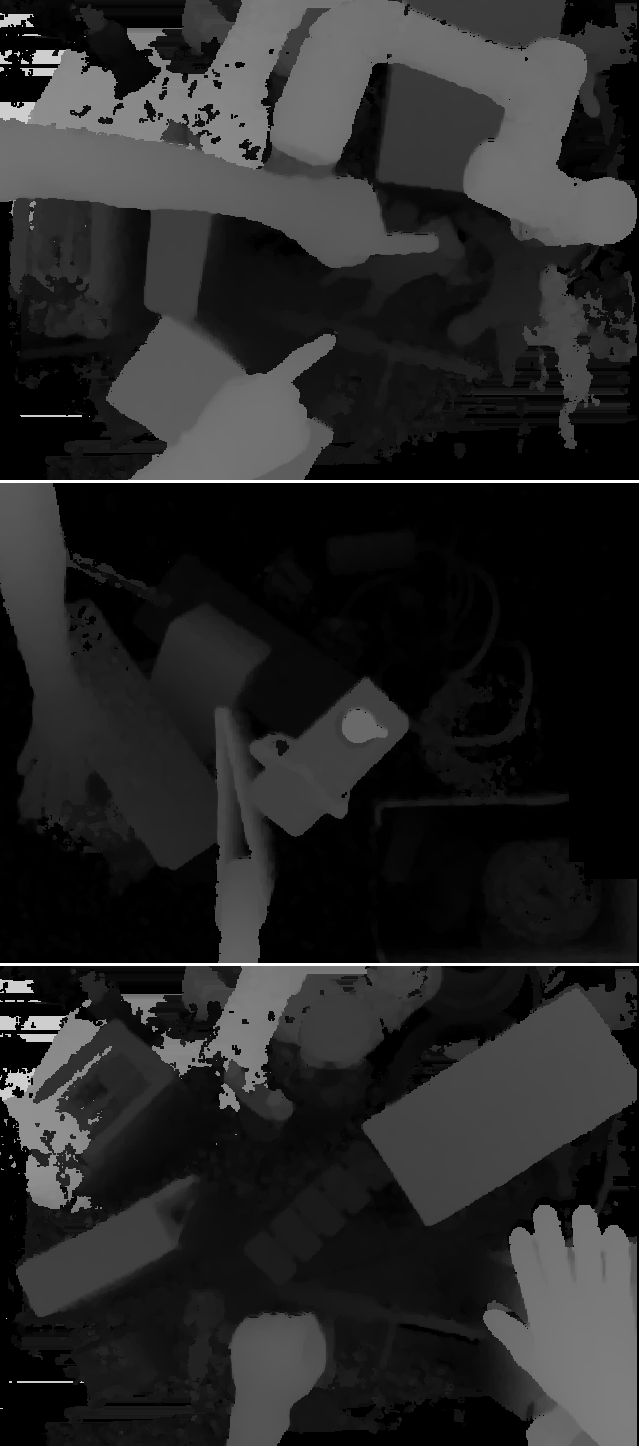}
    \caption{}
\end{subfigure}
\begin{subfigure}[t]{0.2\textwidth}
    \centering
    \includegraphics[height=7.5cm]{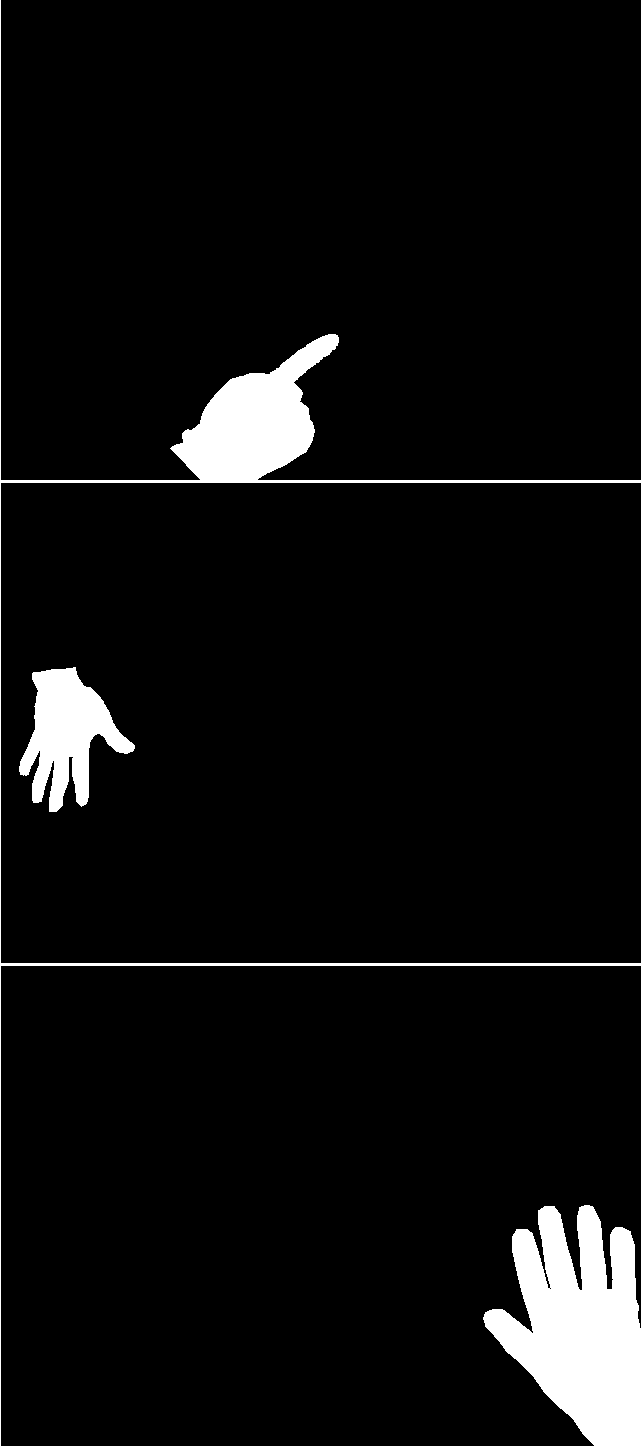}
    \caption{}
\end{subfigure}
\begin{subfigure}[t]{0.2\textwidth}
    \centering
    \includegraphics[height=7.5cm]{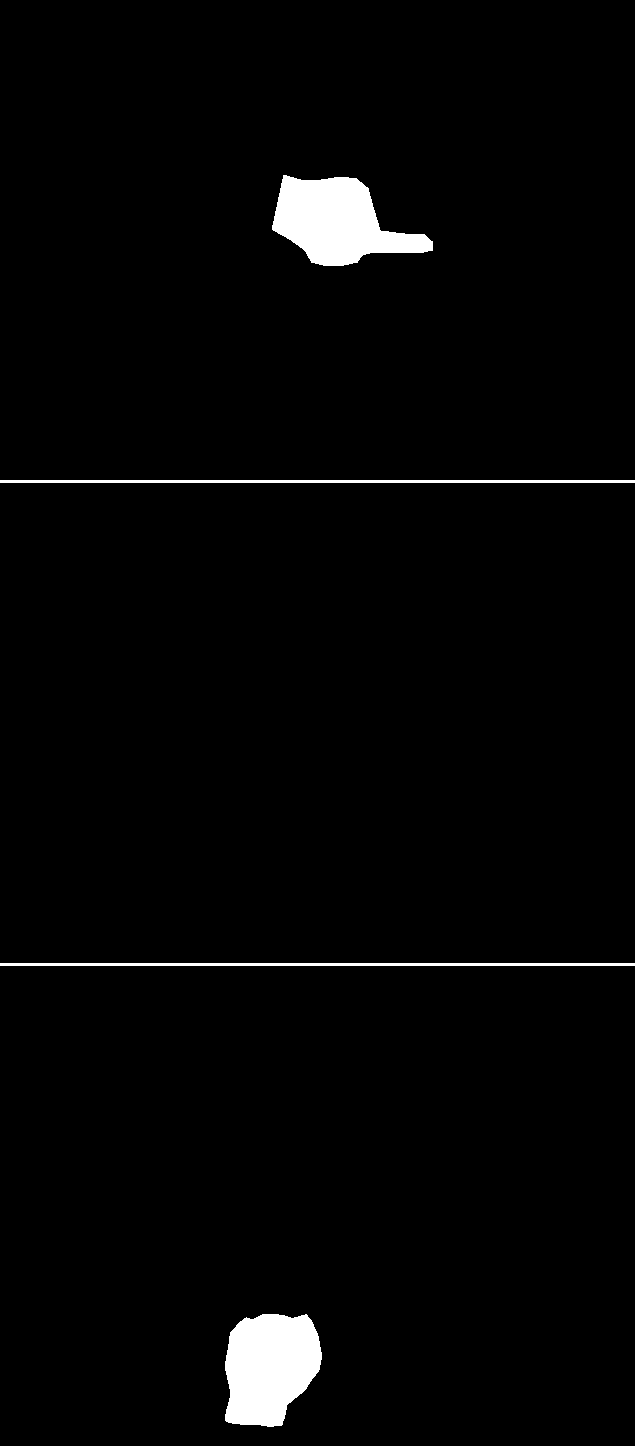}
    \caption{}
\end{subfigure}

	\caption{Real dataset for evaluation: (a) RGB; (b) Depth; (c) mask of the first instance; (d) mask of the second instance.}
	\label{fig_eval_dataset}
\end{figure}

To assess the performance of domain randomization (DR), we compare the results of instance segmentation models trained on different datasets and evaluated on a test dataset corresponding to the cluttered and unstructured industrial environment, which was acquired using a RealSense L515 calibrated to spatially align the depth and RGB frames. During our experiment, we employed a 640x480 depth stream which underwent processing through a modified Intel RealSense librealsense library equipped with improved depth image filtering. To preprocess the data, we utilized a colourization tool that rescaled the depth information (in the range [0.2, 1.0] m to an 8-bit value. Additionally, we utilized a customized hole-filling filter to eliminate depth image shadows that arise from the camera's stereoscopic technology. To restore the missing depth information in the obstacle-free areas, we incorporated a static image of the scene captured in the workplace. Our test dataset contains 706 images (see Figure \ref{fig_eval_dataset}) obtained in different circumstances including variable lighting, background and obstacles, number of hands, different work gloves (red, green, white, yellow) with different lengths of sleeves. For this dataset, the same assumption was made that the system would have only one user, so the maximum number of instances per sample was limited to 2.

\begin{figure}
	\centering
 \begin{subfigure}[t]{0.32\textwidth}
	\centering
    \includegraphics[width=\textwidth]{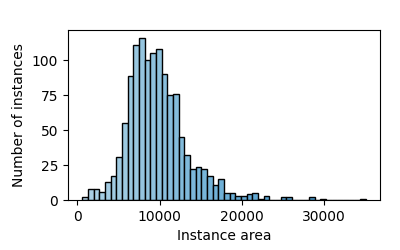}
    \caption{}
\end{subfigure}
\begin{subfigure}[t]{0.32\textwidth}
	\centering
    \includegraphics[width=\textwidth]{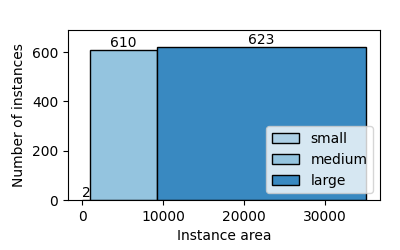}
    \caption{}
\end{subfigure}
\begin{subfigure}[t]{0.32\textwidth}
	\centering
    \includegraphics[width=\textwidth]{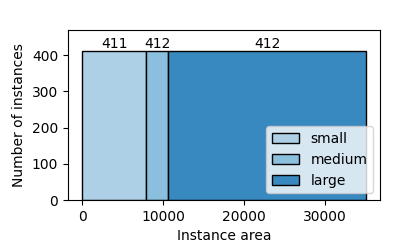}
    \caption{}
\end{subfigure}

    \caption{Object count histogram: (a) instances in our DR dataset; (b) by categorization into small, medium and large object groups according to standard COCO-calculated thresholds – the instances are distributed into categories unevenly; (b) by categorization into small, medium and large object groups according to thresholds calculated for our dataset – instances are distributed evenly into the categories. }
    \label{fig_cat_hist}
\end{figure}

In order to evaluate the average precision for small, medium and large objects separately, we assessed the instances’ area distribution in our dataset (see Figure \ref{fig_cat_hist}a), as the commonly used values from the COCO dataset are clearly out of place for our case, as it is shown in Figure \ref{fig_cat_hist}b, where the “small” category contains only two instances. Figure \ref{fig_cat_hist}c shows the adapted categorization of the instances into small, medium, and large categories according to their sizes. This ensures that each category contains the same number of instances allowing for a fairer comparison.

We hypothesize that to obtain reliable predictions in a complex unstructured environment, it is necessary to employ a multimodal input that integrates both colour and depth information, with each modality contributing to enhancing the accuracy of model predictions. To assess this hypothesis, we investigate the individual impact of each modality, as well as their synergy, on the overall model performance in relation to chosen metrics.

\section{Results}
In this section, we evaluate our approach in a set of experiments which compare the results of the models trained on our DR dataset with the results of the same networks trained on publicly available hands datasets and with existing ready-to-use hand tracking solution MediaPipe Hands \cite{zhang_mediapipe_2020}. The experiments conducted in this study involve training the model exclusively on synthetic datasets without any additional fine-tuning on real images. The purpose of this approach is to evaluate the effectiveness of the proposed method in terms of generalization. The evaluation is done using AP and PDQ metrics and qualitative analysis of the ouputs, where actual predictions are examined.

\subsection{Comparison with COCO-pretrained models}
\begin{figure}[hb]
    \centering
\begin{subfigure}[t]{0.49\textwidth}
    \centering
    \includegraphics[width=\textwidth]{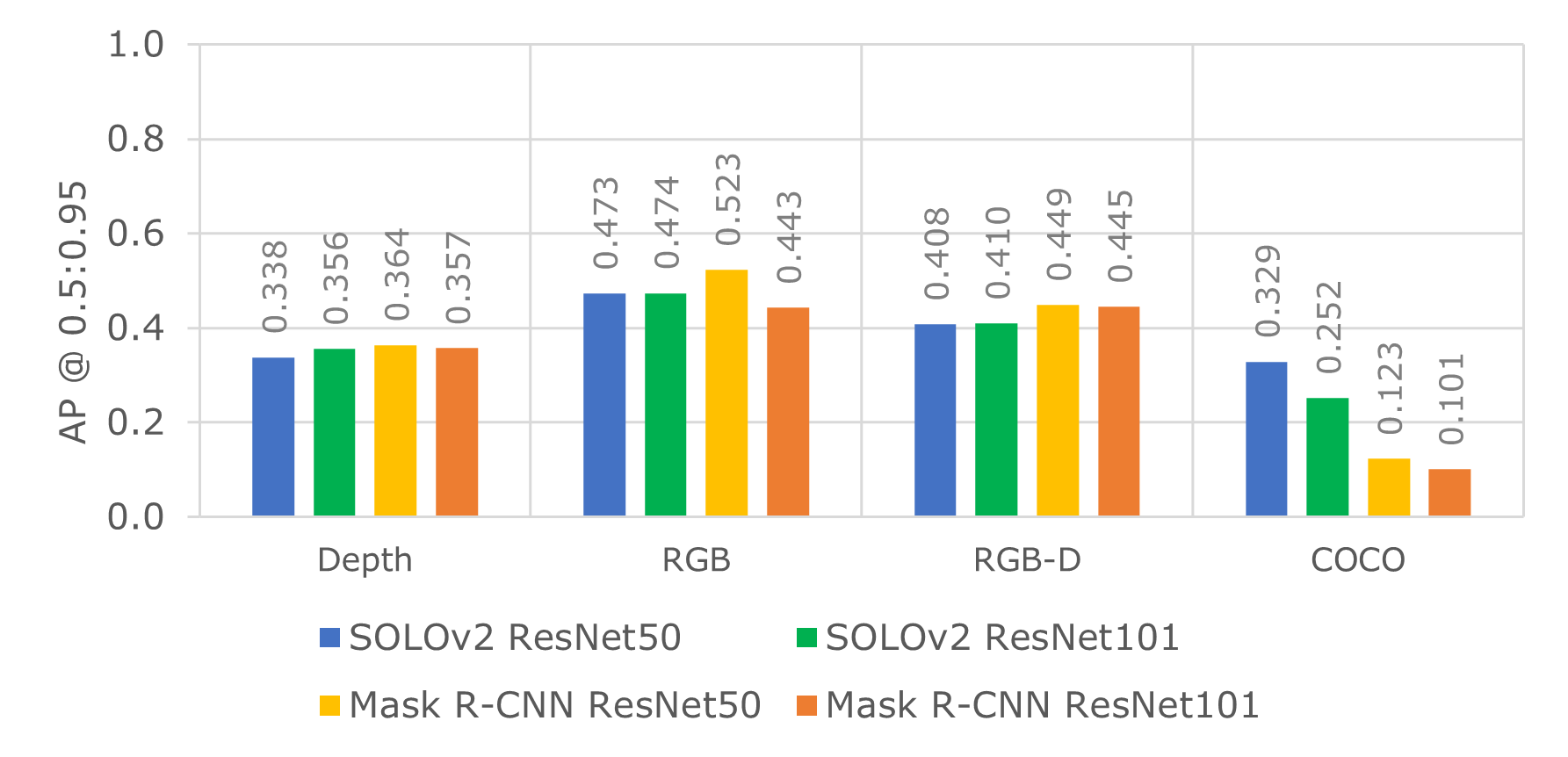}
    \caption{}
\end{subfigure}
\hfill
\begin{subfigure}[t]{0.49\textwidth}
    \centering
    \includegraphics[width=\textwidth]{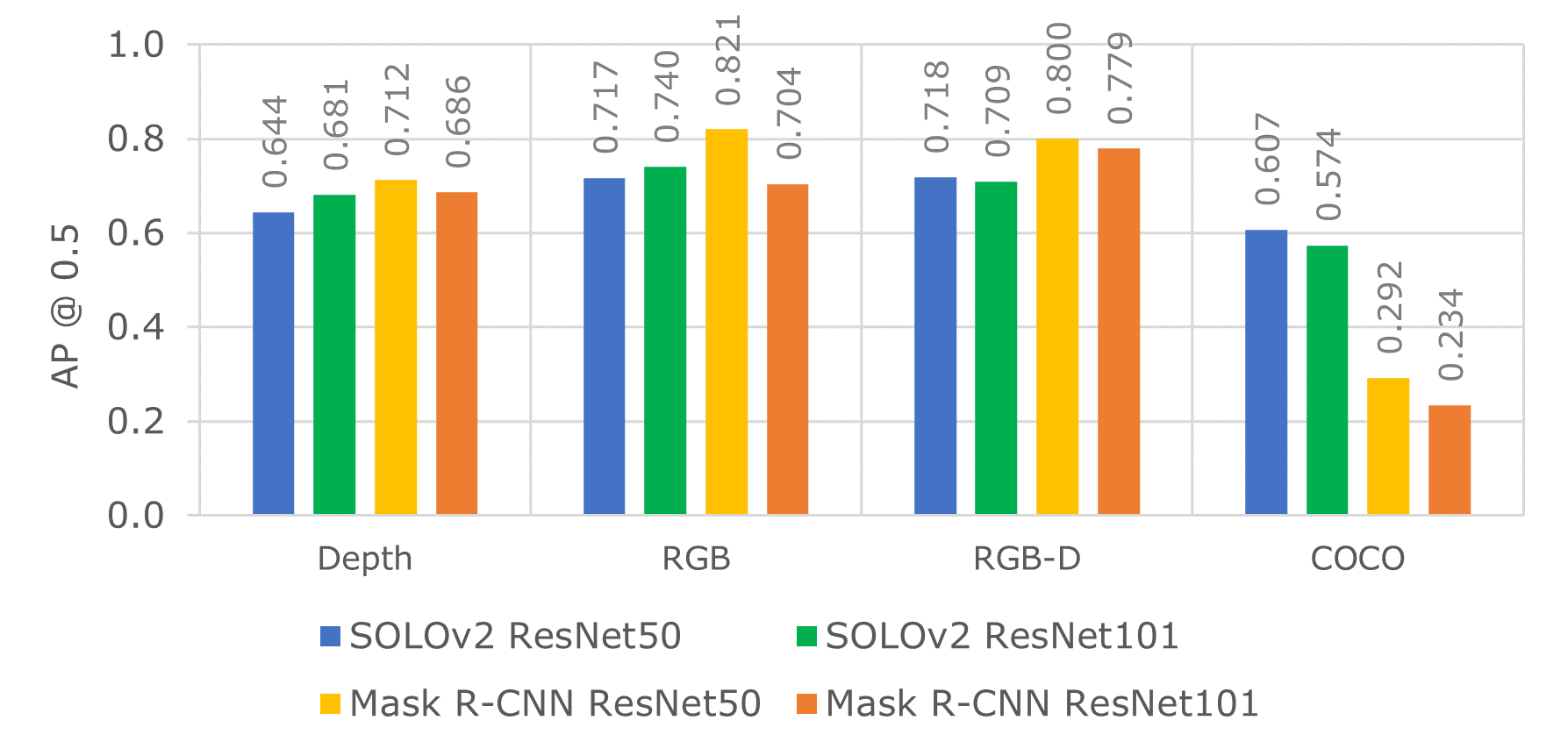}
    \caption{}
\end{subfigure}
    \caption{AP@0.5:0.95 (a) and AP@0.5 (b) evaluated with confidence score threshold 0.1 for different models trained on different modalities: Depth-only, RGB, RGB-D in comparison with a model trained on COCO dataset (Person class only)}
    \label{fig_ap}
\end{figure}

As the first evaluation strategy, we compare the AP for the custom-trained with pretrained RGB COCO models, since COCO dataset itself contains the Person class which indicated a simple possibility for extracting information about the hands. To make a fair comparison with the COCO-pretrained models, a second version of the real dataset was prepared where the masks were adjusted to represent the entire arms. Since IoU is a relative metric, this procedure allows for a fairer assessment of the performance of the models trained with COCO. We evaluate the trained models on the presented test dataset using AP@0.5:0.95 and AP@0.5 (since our task comprises only of a single class) with the default minimum score threshold of 0.1, the results are presented in Table \ref{tab_dataset_stats2} and Figure \ref{fig_ap}. 

\begin{table}
\caption{Average precision and average recall of models trained on our DR dataset with all modalities (Depth, RGB, RGB-D) compared with metrics for COCO. Evaluated on test dataset (real camera images) with minimum confidence score threshold of 0.1. Mask R-CNN ResNet50 achieves highest AP}\label{tab_dataset_stats2}
\begin{tabular}{lcccccc} 
\toprule
Model & 
\begin{tabular}{p{12mm}}
AP @0.5:0.95
\end{tabular}
& 
\begin{tabular}{p{12mm}}
AP@0.5
\end{tabular}
&
\begin{tabular}{p{12mm}} 
AP\textsubscript{small} @0.5:0.95 
\end{tabular}
& 
\begin{tabular}{p{12mm}}
AP\textsubscript{medium} @0.5:0.95 
\end{tabular}
& \begin{tabular}{p{12mm}} 
AP\textsubscript{large} @0.5:0.95 
\end{tabular}
& 
\begin{tabular}{p{12mm}}
AR @0.5:0.95
\end{tabular}\\
\midrule

SOLOv2 ResNet50 (Depth) & 0.338 & 0.644 & 0.263 & 0.369 & 0.405 & 0.265 \\
SOLOv2 ResNet101 (Depth) & 0.356 & 0.681 & 0.293 & 0.392 & \textbf{0.415} & \textbf{0.274} \\
Mask R-CNN ResNet50 (Depth) & \textbf{0.364} & \textbf{0.712} & 0.291 & \textbf{0.424} & \textbf{0.415} & 0.273 \\
Mask R-CNN ResNet101 (Depth) & 0.357 & 0.686 & \textbf{0.308} & 0.415 & 0.394 & \textbf{0.274} \\
\midrule
SOLOv2 ResNet50 (RGB) & 0.473 & 0.717 & 0.399 & 0.578 & 0.557 & 0.330 \\
SOLOv2 ResNet101 (RGB) & 0.474 & 0.740 & 0.391 & 0.583 & 0.550 & 0.336 \\
Mask R-CNN ResNet50 (RGB) & \textbf{0.523} & \textbf{0.821} & \textbf{0.459} & \textbf{0.612} & \textbf{0.576} & \textbf{0.357} \\
Mask R-CNN ResNet101 (RGB) & 0.443 & 0.704 & 0.388 & 0.567 & 0.497 & 0.321 \\
\midrule
SOLOv2 ResNet50 (RGB-D) & 0.408 & 0.718 & 0.326 & 0.465 & 0.460 & 0.306 \\
SOLOv2 ResNet101 (RGB-D) & 0.410 & 0.709 & 0.348 & 0.457 & 0.458 & 0.306 \\
Mask R-CNN ResNet50 (RGB-D) & \textbf{0.449} & \textbf{0.800} & 0.412 & \textbf{0.505} & \textbf{0.479} & \textbf{0.325} \\
Mask R-CNN ResNet101 (RGB-D) & 0.445 & 0.779 & \textbf{0.423} & 0.504 & 0.468 & 0.323 \\
\midrule
SOLOv2 ResNet50 (COCO) & \textbf{0.329} & \textbf{0.607} & \textbf{0.394} & \textbf{0.490} & \textbf{0.274} & \textbf{0.281} \\
SOLOv2 ResNet101 (COCO) & 0.252 & 0.574 & 0.278 & 0.349 & 0.244 & 0.225 \\
Mask R-CNN ResNet50 (COCO) & 0.123 & 0.292 & 0.218 & 0.322 & 0.082 & 0.157 \\
Mask R-CNN ResNet101 (COCO) & 0.101 & 0.234 & 0.221 & 0.284 & 0.082 & 0.148 \\
\bottomrule
\end{tabular}
\end{table}

\begin{figure}
	\centering
		\includegraphics[width=14cm]{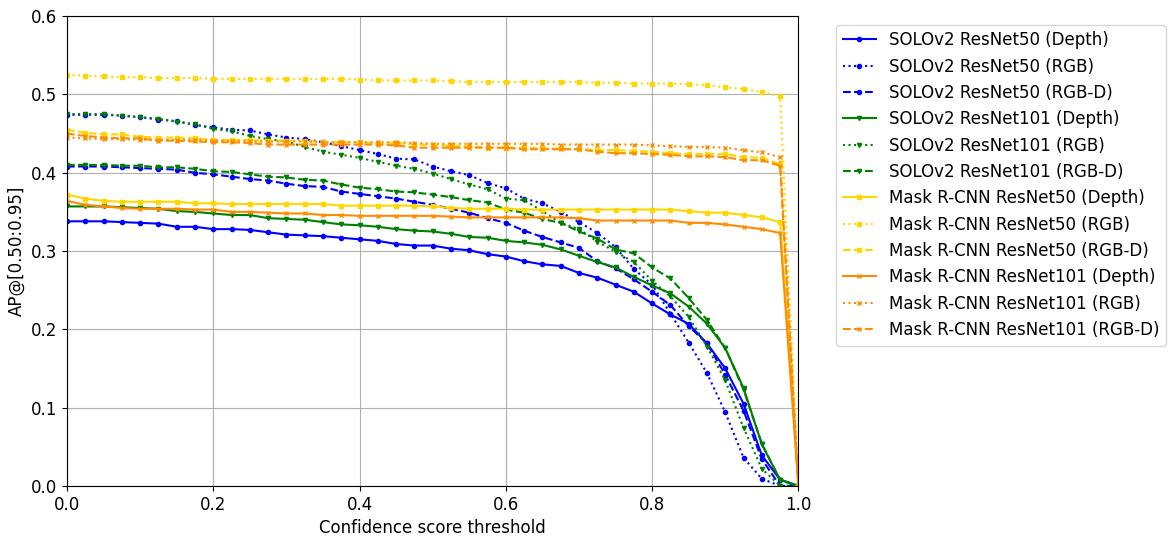}
	\caption{AP curves (higher is better) for trained models evaluated for different confidence score thresholds. Mask R-CNN models retain their AP up to a confidence threshold of 0.95.}
	\label{fig_apscore}
\end{figure}

It can be noticed that, in general, the Mask R-CNN models have better results for all modalities (the only exception is for the pretrained COCO RGB models, where the results of the models with the corresponding backbones differ by a factor of ~2.5). In terms of modalities, RGB enabled the best results for all models, while the RGB-D models performed comparably or worse by up to 6.5 AP, achieving AP 44.9 for Mask R-CNN ResNet50 model. The depth models performed similarly and significantly worse than the RGB models, reaching a maximum of 36.4 AP on Mask R-CNN with the ResNet50 backbone. We assume that these results are due to differences in the properties of the simulated and real depth images. The best AP was obtained with the Mask R-CNN ResNet50 RGB, which reached an AP of 52.3. Ratios of the results of the models evaluated using AP@0.5 are similar, however, the differences between their performances are less prominent, especially considering RGB and RGB-D models. Best-performing models remain the same for all modalities – Mask R-CNN, achieving 82.1 AP for RGB input. This outcome is somewhat surprising, as it was anticipated that the use of RGB-D data, which provides more information, would result in improved predictions compared to RGB.
All models trained on our DR dataset show a significant increase in metrics compared to models trained using COCO dataset. We assume that the lower AP shown by models trained with COCO are due to the fact that these models perform better when a larger portion of the human body can be observed in the image. When testing with a real camera, we noticed that the pretrained COCO models with Person class performed better when the image contained part of the human body, however, if only the hands were visible or if gloves were on the palms, the results deteriorated significantly.
It was also found that the SOLO models had significantly lower prediction confidence scores compared to the corresponding Mask R-CNN models. This is partly due to the way the confidence scores are calculated in each approach, however, we assume that the differences in the characteristics of the generated DR dataset and the actual camera images also play a role in the low prediction scores. Evaluation of the models using the COCO evaluation API for different confidence score thresholds leads to similar results for all models (see Figure \ref{fig_apscore}). The AP of the models monotonically decreases with increasing thresholds, but the trend of the decrease is quite different for the Mask R-CNN and SOLOv2 models, where the deterioration of the results is much steeper for SOLOv2, while Mask R-CNN maintains a high AP up to a reliability score threshold of 0.95 without any significant deterioration. The other COCO metrics (AP\textsubscript{small}, AP\textsubscript{medium}, AP\textsubscript{large}, etc.) show similar trends for each model, but the ratios between them are ordered differently for each model than for the AP comparison. 

The choice of the confidence score threshold plays a critical role in evaluating the performance of a model. Lower confidence thresholds may result in a higher number of false positives, while higher thresholds result in more false negatives. The mean average precision score, as used in the COCO evaluation, penalizes false positives only to a marginal extent \cite{hall_probabilistic_2020}: multiple  true-positive bounding boxes predicted for a single instance are not penalized as false positives, as long as they satisfy the IoU threshold used in the evaluation. However, in practical applications, a high number of false positives can have significant negative consequences (especially in the context of robotics).

\begin{figure}
	\centering
		\includegraphics[width=14cm]{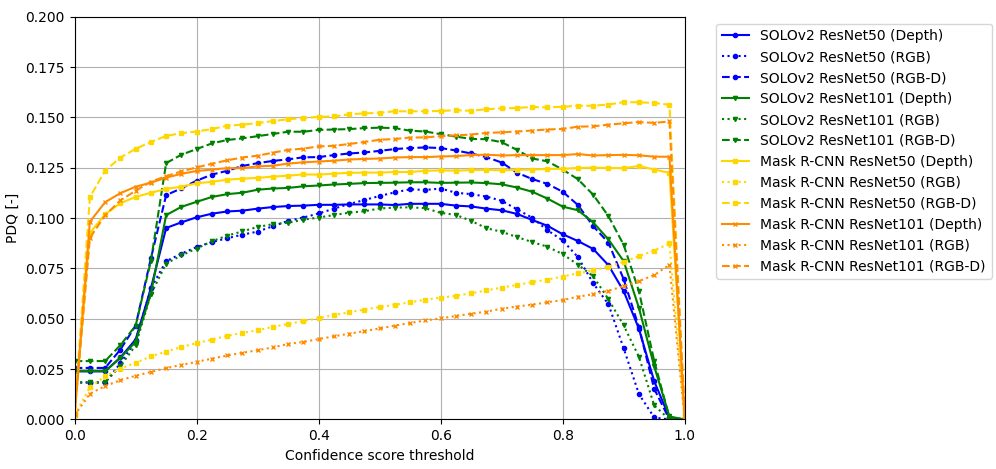}
	\caption{PDQ curves (higher is better) obtained by evaluating PDQ as a function of confidence score threshold. Evaluated on test dataset (bounding boxes).}
	\label{fig_pdqcurves}
\end{figure}

To further investigate the performance of our models we utilize probability-based detection quality (PDQ) metric presented by Hall et al. \cite{hall_probabilistic_2020}. PDQ deals with a probabilistic evaluation of object detection, explicitly penalizing false positives and false negatives. PDQ evaluates the best match between the detected instances and ground-truth labels for each sample, with mismatched  bounding boxes considered false positives. Search for optimal confidence score threshold was performed by evaluating PDQ as a function of confidence score threshold in the range [0.0,1.0] with 0.025 steps (see Figure \ref{fig_pdqcurves}), following the experiment presented by Wenkel et al. \cite{wenkel_confidence_2021}. The PDQ evaluation currently supports only the bounding box-based evaluation, thus it was performed using the bounding boxes predicted by the Mask R-CNN models and bounding boxes calculated for the masks predicted by the SOLOv2 models.

The calculation of the PDQ as a function of the confidence score threshold results in curves that are distinct from the curves obtained from the COCO metrics (Figure \ref{fig_apscore}). By considering these PDQ curves, we identify the confidence score threshold that represents the optimal operating point for each of the models. Unlike the COCO evaluation, PDQ imposes penalization on both false negatives and false positives while also assessing the confidence scores of the predictions. Therefore, the curve could be interpreted as follows: selecting a confidence score threshold lower than the one corresponding to the maximum PDQ value results in a higher occurrence of false positives while choosing a higher threshold leads to an increasing number of false negatives. Each model’s curve is distinctive, however, in general, Mask R-CNN and SOLOv2 models have their own typical trends. Table \ref{tab_dataset_stats3} presents the findings of the optimal confidence score analysis, whereby the threshold yielding the maximum PDQ is selected as the optimal confidence threshold. Additionally, the table provides information on the AP value at the same confidence threshold, along with the highest AP value attained by the model (which is achieved by all models at the score threshold of 0, as illustrated in Figure \ref{fig_pdqcurves}).

\begin{table}[ht]
\centering
\caption{Summary of PDQ bounding box evaluation, the maximum PDQ is presented along with the corresponding confidence score threshold and AP obtained. PDQ\textsubscript{max} – maximum PDQ obtained by the model; Confidence score threshold at PDQ\textsubscript{max} – represents score threshold at which PDQ\textsubscript{max} was obtained (optimal confidence threshold); AP at PDQ\textsubscript{max} – AP that corresponds to PDQ\textsubscript{max}; AP\textsubscript{max} – maximum AP obtained by the model. }\label{tab_dataset_stats3}
\begin{tabular}{lcccc}
\toprule
Model & PDQ\textsubscript{max} &
\begin{tabular}{p{30mm}}
Confidence score threshold at PDQ\textsubscript{max}
\end{tabular}
& 
\begin{tabular}{p{20mm}}
AP at PDQ\textsubscript{max}
\end{tabular}& AP\textsubscript{max} \\
\midrule
SOLOv2 ResNet50 (Depth) & 0.1071 & 0.550 & 0.301 & 0.338 \\
SOLOv2 ResNet101 (Depth) & 0.1178 & 0.550 & 0.318 & 0.357 \\
Mask R-CNN ResNet50 (Depth) & 0.1259 & 0.925 & 0.346 & \textbf{0.372} \\
Mask R-CNN ResNet101 (Depth) & \textbf{0.1318} & 0.825 & 0.339 & 0.364 \\
\midrule
SOLOv2 ResNet50 (RGB) & \textbf{0.1147} & 0.600 & 0.380 & 0.474 \\
SOLOv2 ResNet101 (RGB) & 0.1054 & 0.550 & 0.385 & 0.475 \\
Mask R-CNN ResNet50 (RGB) & 0.0874 & 0.975 & 0.498 & \textbf{0.525} \\
Mask R-CNN ResNet101 (RGB) & 0.0768 & 0.975 & 0.420 & 0.445 \\
\midrule
SOLOv2 ResNet50 (RGB-D) & 0.1351 & 0.575 & 0.342 & 0.408 \\
SOLOv2 ResNet101 (RGB-D) & 0.1449 & 0.500 & 0.372 & 0.410 \\
Mask R-CNN ResNet50 (RGB-D) & \textbf{0.1576} & 0.900 & 0.424 & \textbf{0.455} \\
Mask R-CNN ResNet101 (RGB-D) & 0.1480 & 0.975 & 0.409 & 0.450 \\
\bottomrule
\end{tabular}
\end{table}

SOLOv2 models typically have optimal confidence score threshold in the range [0.5,0.6], while Mask R-CNN models tend to perform the best with a score threshold over 0.825. According to the results shown in Table \ref{tab_dataset_stats3}, the best-performing model is Mask R-CNN ResNet50 RGB-D obtaining up to 0.1576 PDQ at a confidence score threshold of 0.9, while retaining high AP. 
 
\begin{figure}[ht]
	\centering
\begin{subfigure}[t]{0.13\textwidth}
    \includegraphics[width=\textwidth]{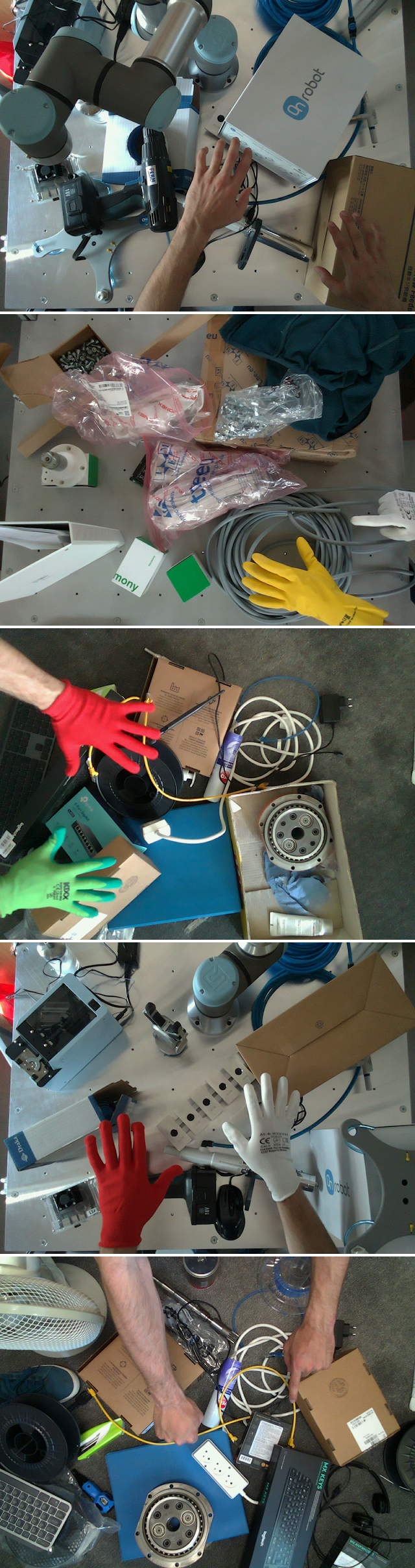}
    \caption{}
\end{subfigure}
\begin{subfigure}[t]{0.13\textwidth}
\centering
    \includegraphics[width=\textwidth]{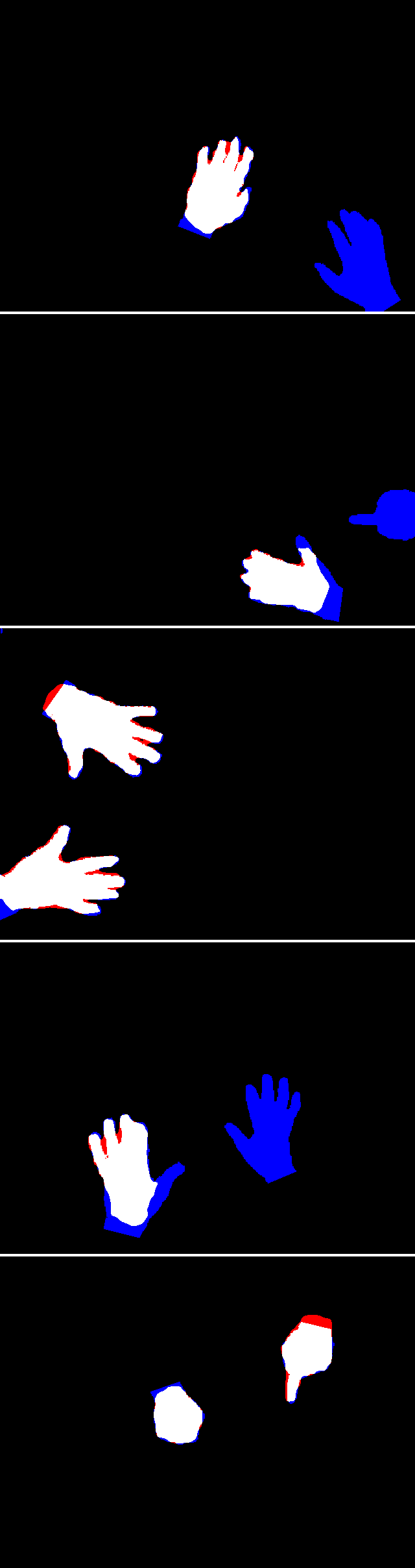}
    \caption{}
\end{subfigure}
\begin{subfigure}[t]{0.13\textwidth}
\centering
    \includegraphics[width=\textwidth]{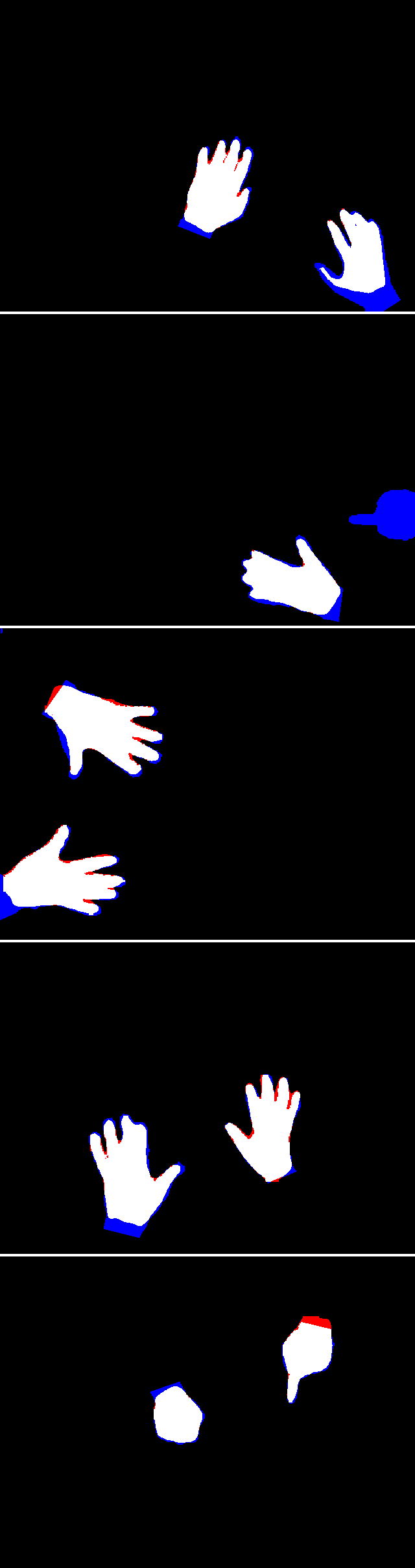}
    \caption{}
\end{subfigure}
\begin{subfigure}[t]{0.13\textwidth}
\centering
    \includegraphics[width=\textwidth]{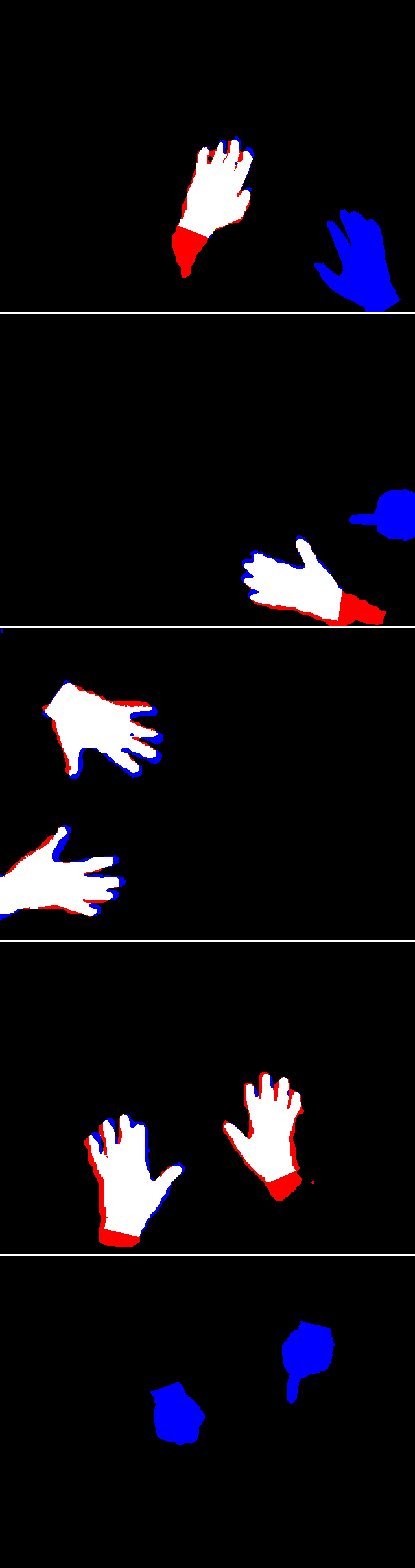}
    \caption{}
\end{subfigure}
\begin{subfigure}[t]{0.13\textwidth}
\centering
    \includegraphics[width=\textwidth]{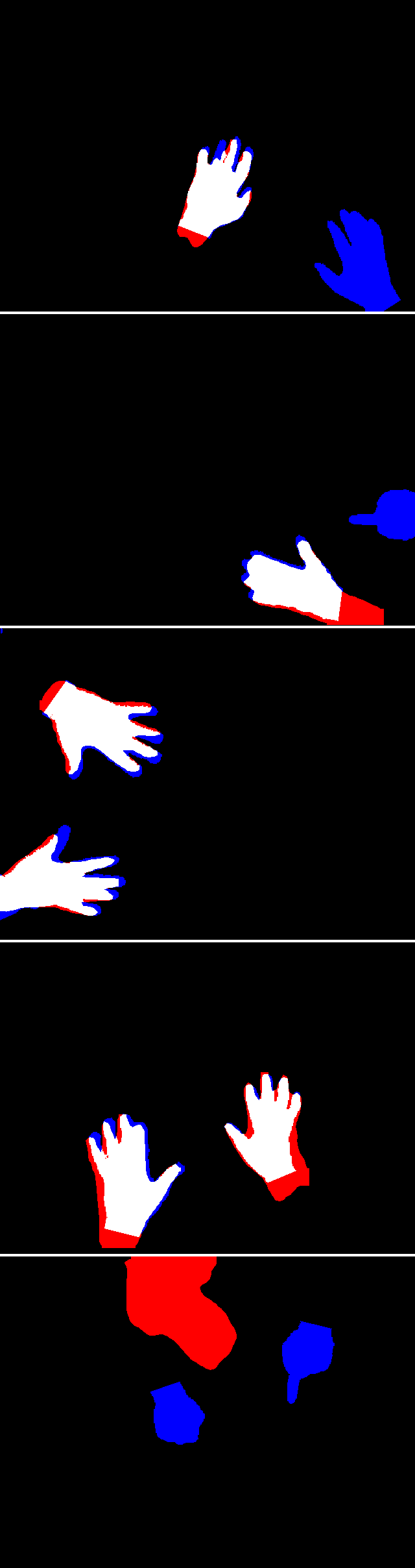}
    \caption{}
\end{subfigure}
\begin{subfigure}[t]{0.13\textwidth}
\centering
    \includegraphics[width=\textwidth]{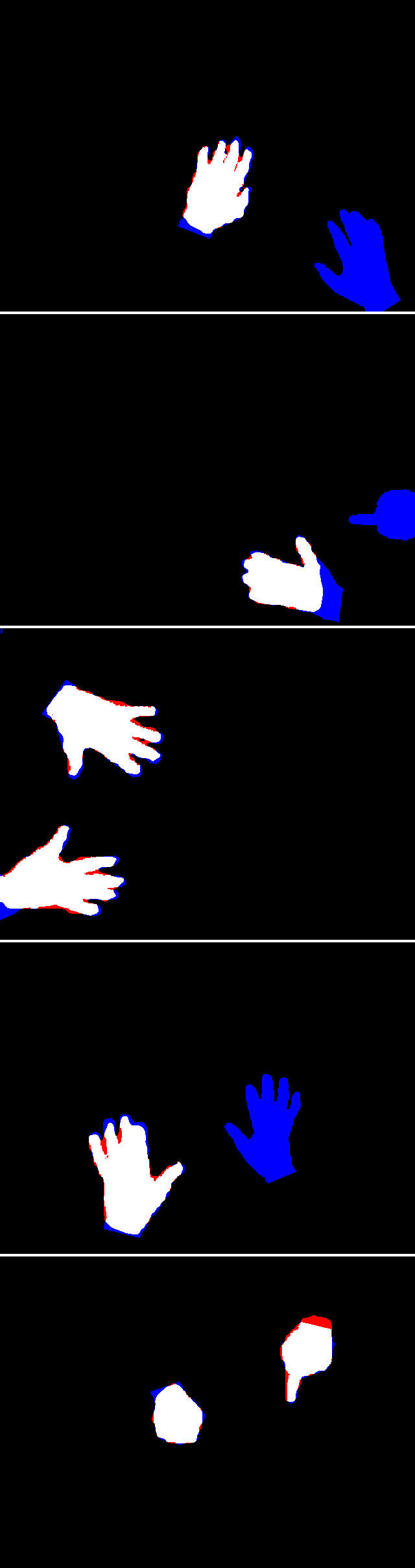}
    \caption{}
\end{subfigure}
\begin{subfigure}[t]{0.13\textwidth}
\centering
    \includegraphics[width=\textwidth]{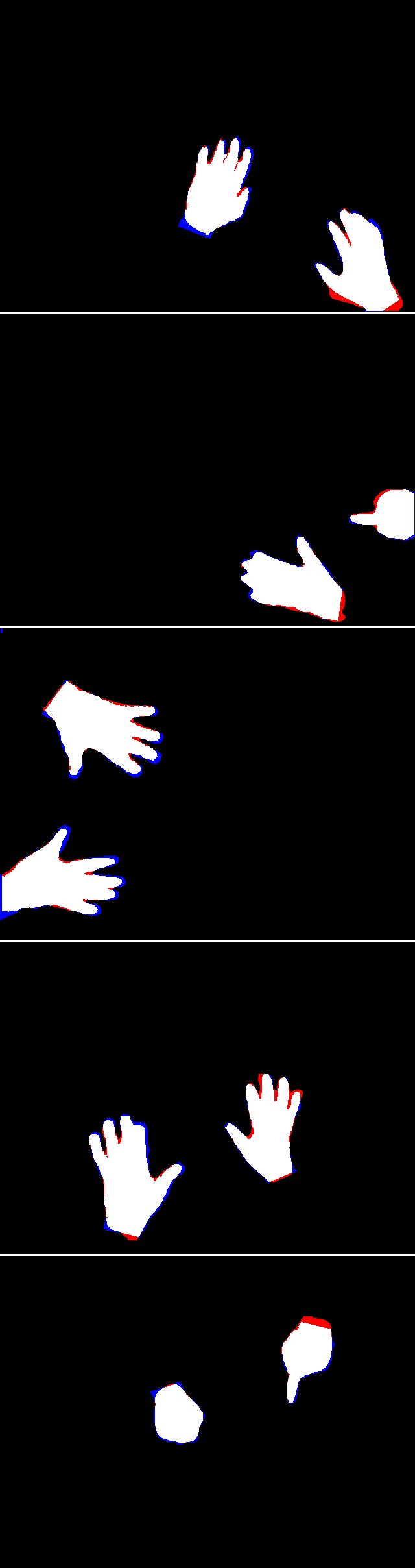}
    \caption{}
\end{subfigure}
    \caption{Qualitative results on test dataset. Model predictions: (a) RGB image of the scene; (b) SOLOv2 ResNet-101 Depth, threshold: 0.55; (c) Mask R-CNN ResNet-101 Depth, threshold: 0.825; (d) SOLOv2 ResNet-50 RGB, threshold: 0.6; (e) Mask R-CNN ResNet-50 RGB, threshold: 0.975; (f) SOLOv2 ResNet-101 RGB-D, threshold: 0.5; (g) Mask R-CNN ResNet-50 RGB-D, threshold: 0.9. True positive (white), false positive (red), and false negative (blue) pixels in the predicted output compared to the ground truth. }
    \label{fig_quality}
\end{figure} 

Apart from that, the RGB-D models performed consistently better in terms of PDQ metric, while the previously high-performing Mask R-CNN RGB models obtained the worst results, whereas the corresponding depth-based models achieved results that were up to two times better.

Figure \ref{fig_quality} shows several sample predictions for the evaluated models. The examples include various environments, lighting conditions as well as different working gloves, and combinations of backgrounds which makes instance segmentation challenging. 

The results observed during qualitative evaluation additionally support the conclusions drawn by PDQ evaluation. The qualitative evaluation was performed by comparing the mask generated by the model with the mask (ground truth) of the image from the test dataset. The evaluation threshold was set according to the previous PDQ analysis, due to this setting the predictions do not contain false positive regions to a large extent. Compared to the AP evaluation, where the models working with RGB input showed the best results, in the qualitative evaluation these models achieved the worst result in terms of the quality of the identification of the instances. In columns Figure \ref{fig_quality}d and Figure \ref{fig_quality}e, corresponding to RGB images, 6 out of 10 instances are identified, in columns Figure \ref{fig_quality}a and Figure \ref{fig_quality}f, corresponding to the SOLOv2 models identified 7 out of 10 instances when using depth-only and RGB-D inputs. The best results are achieved by the models trained by Mask R-CNN models, where 9 out of 10 instances are identified when using depth Figure \ref{fig_quality}b and all instances are identified with RGB-D input Figure \ref{fig_quality}g. Prediction based on RGB-D images according to the qualitative test shows the best results in terms of the number and quality of instance detection. From the point of view of the difficulty of recognizing the hand instance, the cases of using a white glove and under the conditions of background and hand colour blending were the most difficult; here the RGB-D models showed the best results, while RGB models contained a higher number of false positives. In contrast, instances where the hand colour was significantly different from the background were identified well by all models. 

\subsection{Comparison with existing datasets}\label{comparison_with_existing_datasets}

\begin{figure}[ht]
	\centering
\begin{subfigure}[t]{0.3\textwidth}
\centering
    \includegraphics[width=\textwidth]{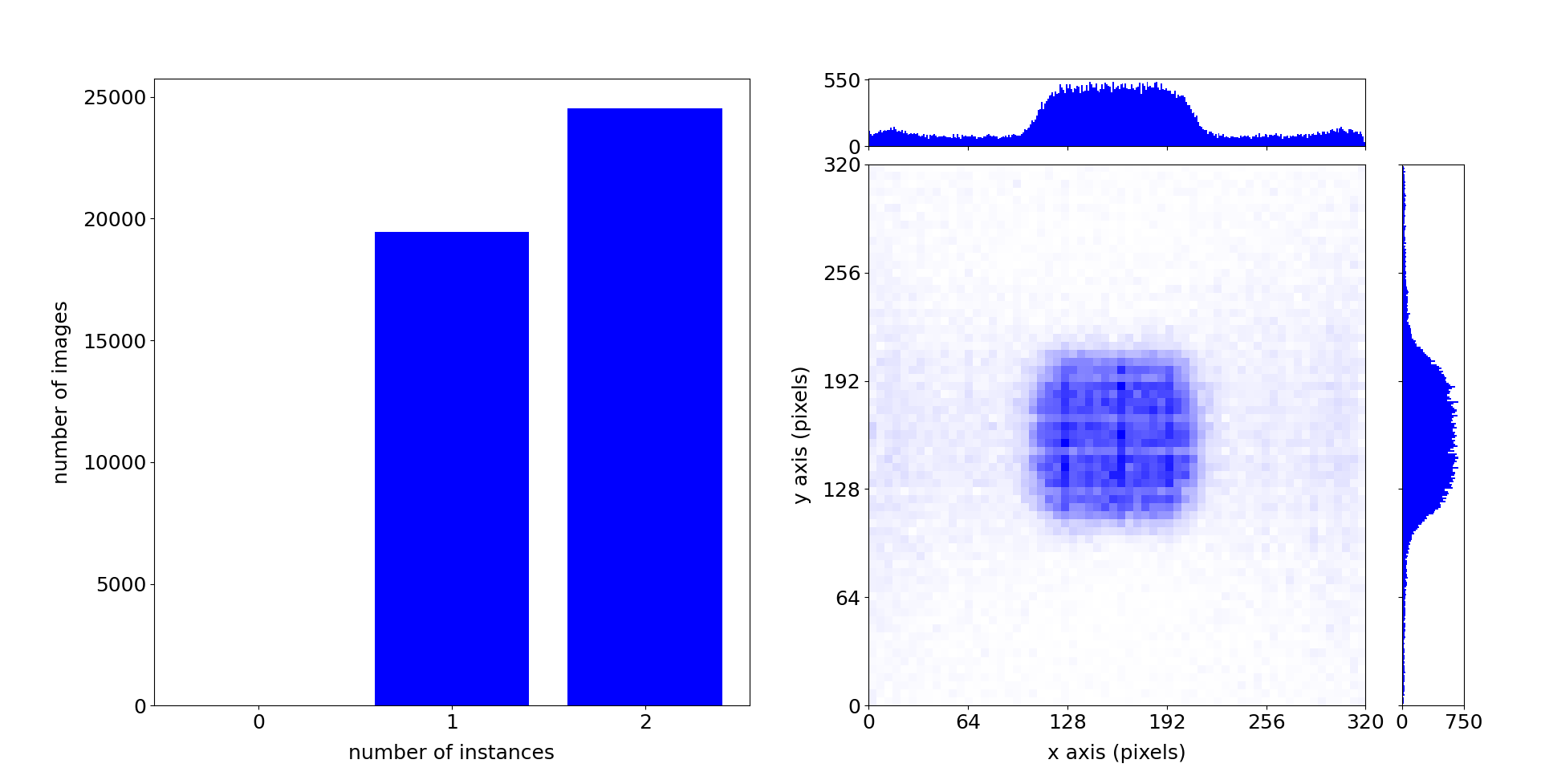}
    \caption{}
\end{subfigure}
\begin{subfigure}[t]{0.3\textwidth}
\centering
    \includegraphics[width=\textwidth]{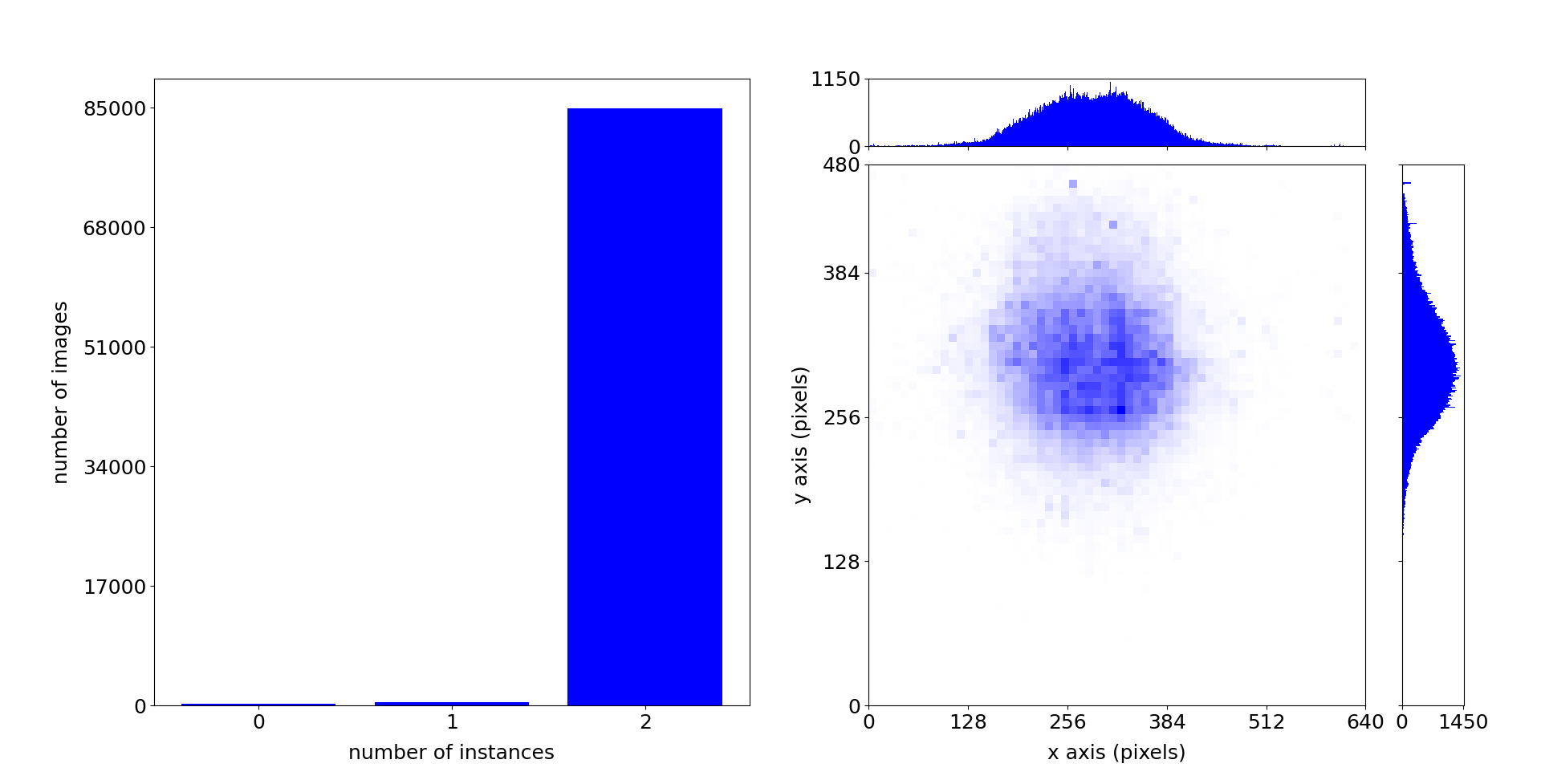}
    \caption{}
\end{subfigure}
\begin{subfigure}[t]{0.3\textwidth}
\centering
    \includegraphics[width=\textwidth]{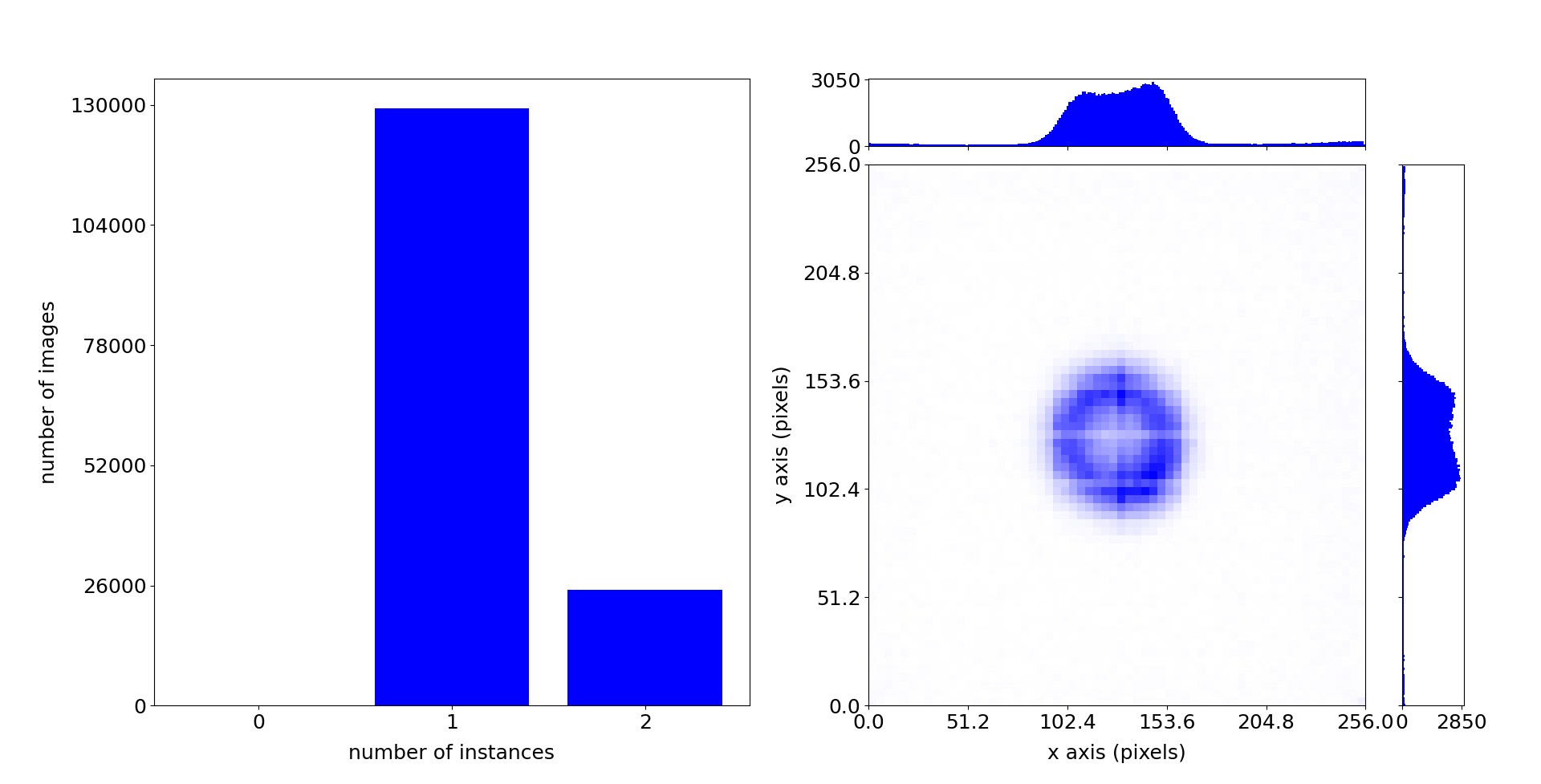}
    \caption{}
\end{subfigure}
\begin{subfigure}[t]{0.3\textwidth}
\centering
    \includegraphics[width=\textwidth]{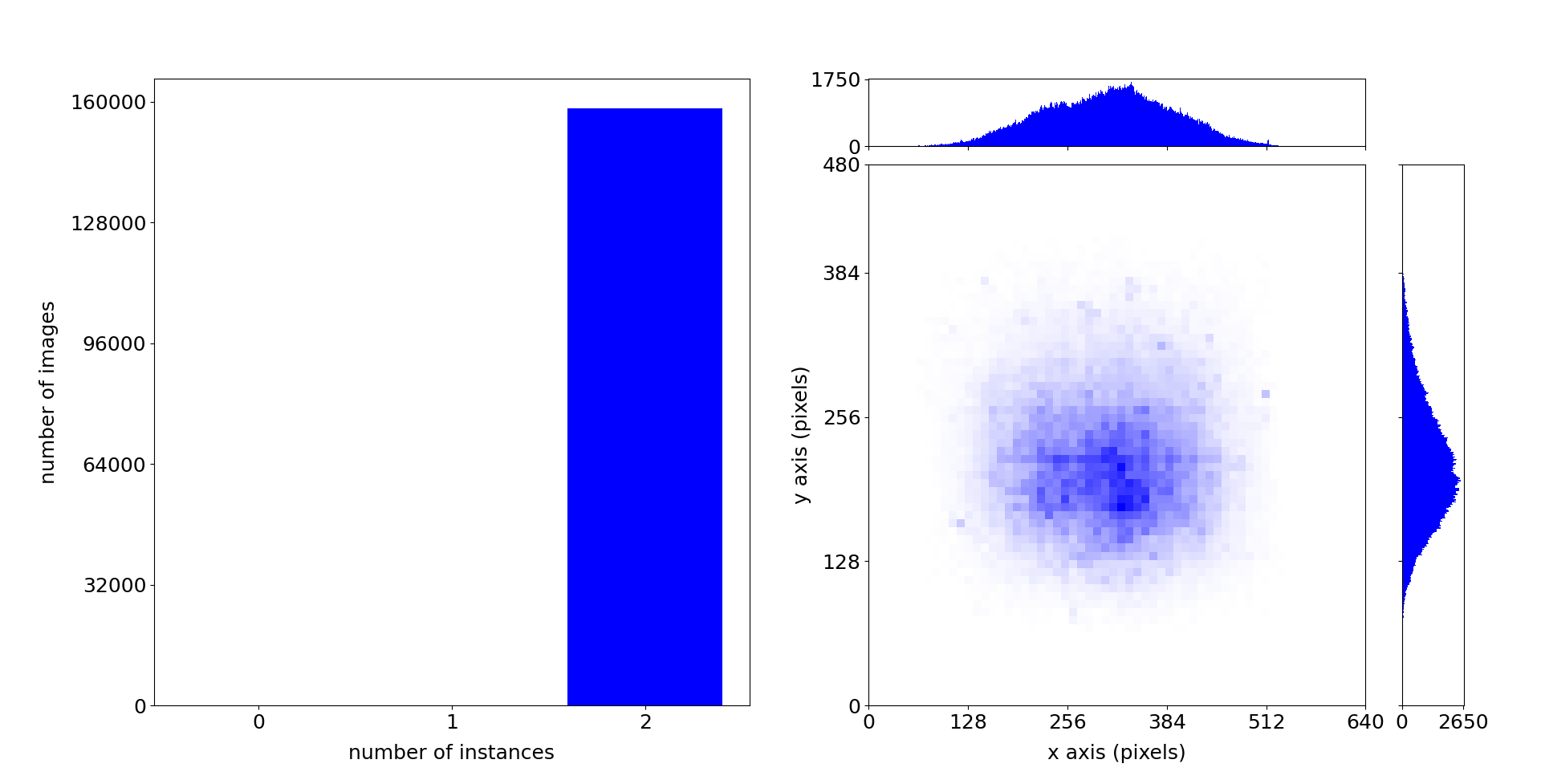}
    \caption{}
\end{subfigure}
\begin{subfigure}[t]{0.3\textwidth}
\centering
    \includegraphics[width=\textwidth]{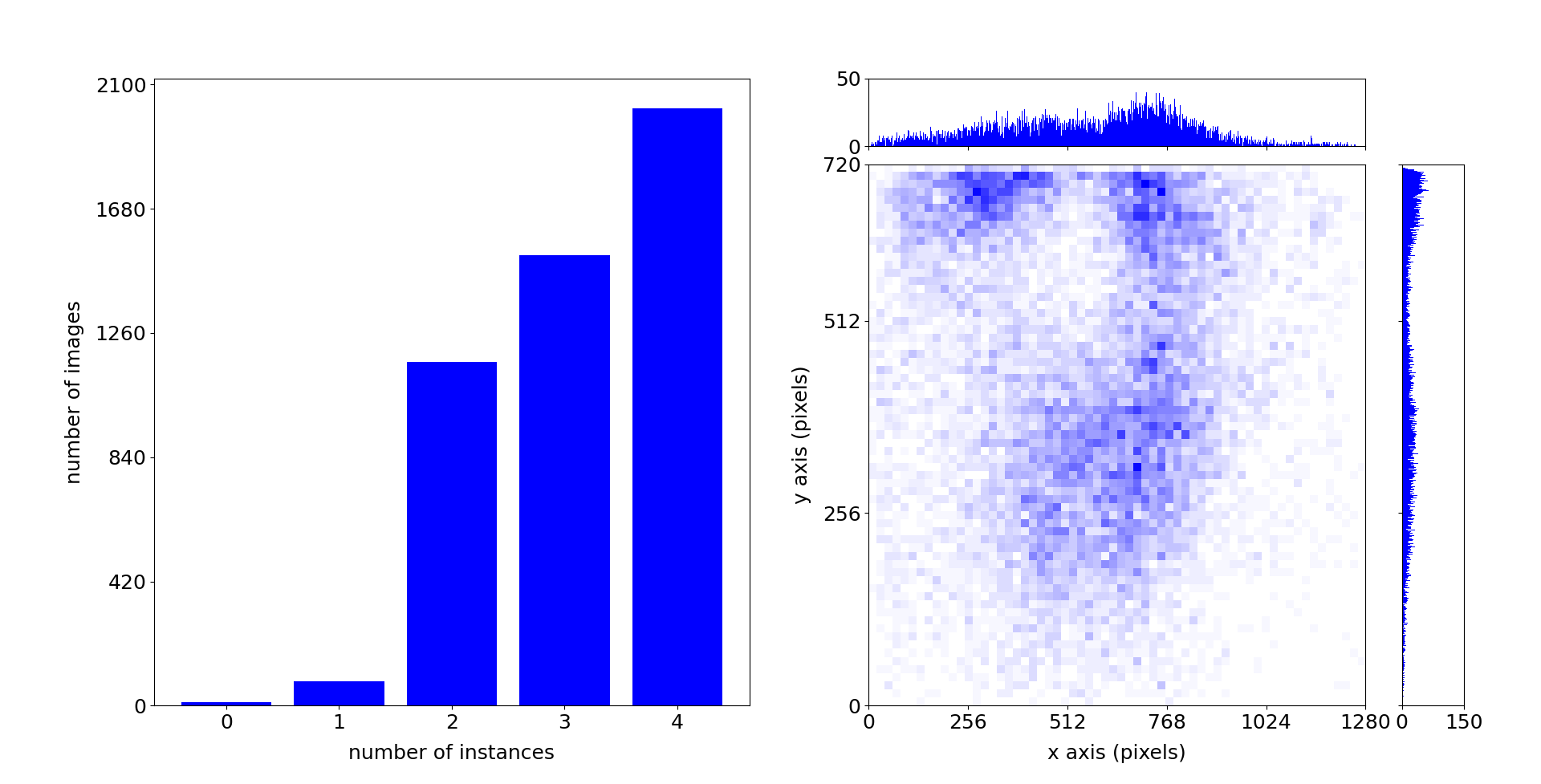}
    \caption{}
\end{subfigure}
    \caption{Statistics of the datasets: (a) RHD; (b) Dense Hands; (c) ObMan; (d) HandSeg; (e) EgoHands. Graphs depict distributions of the number of instances per image (left) and instance centroid locations (right). Applying DR simplifies the generation of data with a wider distribution (see Figure \ref{fig_statistics_our}).}
    \label{fig_stats}
\end{figure}

To benchmark our dataset against existing works in this field, we adapted several popular publicly available hand datasets (see Table \ref{tab_dataset_stats4}, descriptions are provided in Relate work section). Statistical distributions of these datasets are shown in Figure \ref{fig_stats} - it can be observed that instances are biased towards the centres of the images. In order to adapt the datasets to our pipeline, the following modifications were applied:

\begin{itemize}
    \item \textbf{Unified and merged class masks:} DenseHands - dense correspondence maps were binarized and used as masks for hand instances; ObMan and RHD  - all masks except hands were omitted.
\item \textbf{Depth range:} The [0.2,1.0] m depth range was mapped to the byte range according to the settings in the test environment. The remaining range was truncated to the 1 m boundary.

\end{itemize}

\begin{table}[hb]

\caption{Existing hand datasets – comparison of the features}\label{tab_dataset_stats4}
\centering
\begin{tabular}{lcccc}
\toprule
Dataset & Annotation method & Samples source & Instances & Modalities \\
\midrule
EgoHands & Manual & Real & Up to 4 & RGB \\
HandSeg & Automatic (marker/gloves) & Real & Up to 2 & RGB-D \\
DenseHands & Automatic & Synthetic & Up to 2 & Depth \\
Rendered Hand Pose (RHD) & Automatic & Synthetic & Up to 2 & RGB-D \\
ObMan & Automatic & Synthetic & Up to 2 & RGB-D \\
\textbf{HaDR (Ours)} & Automatic & Synthetic & Up to 2 & RGB-D \\
\bottomrule
\end{tabular}
\end{table}

\begin{table}[ht]
    \caption{AP and PDQ evaluations for models trained on existing datasets (evaluated on test dataset)}\label{tab_dataset_stats5}
    \centering
    \begin{tabular}{lcccc}
        \toprule
        Model
        & 
        \begin{tabular}{p{10mm}}
        PDQ\textsubscript{max}
        \end{tabular}
        & 
        \begin{tabular}{p{25mm}}
        Confidence score threshold at PDQ\textsubscript{max}
        \end{tabular}
        & AP at PDQ\textsubscript{max}
        & AP\textsubscript{max} \\
        
        \midrule
        
        DenseHands: SOLOv2 ResNet50 (Depth) & 0.0000 & 0.000 & 0.9 & 0.9 \\
        DenseHands: SOLOv2 ResNet101 (Depth) & 0.0001 & 0.225 & 0.8 & 0.8 \\
        DenseHands: Mask R-CNN ResNet50 (Depth) & \textbf{0.0061} & 0.975 & 8.4 & \textbf{9.3} \\
        DenseHands: Mask R-CNN ResNet101 (Depth) & 0.0034 & 0.975 & 2.7 & 2.9 \\
        \midrule
        HandSeg: SOLOv2 ResNet50 (Depth) & 0.0104 & 0.350 & 3.5 & 3.9 \\
        HandSeg: SOLOv2 ResNet101 (Depth) & \textbf{0.0194} & 0.450 & 6.0 & \textbf{6.9} \\
        HandSeg: Mask R-CNN ResNet50 (Depth) & 0.0147 & 0.975 & 1.8 & 2.0 \\
        HandSeg: Mask R-CNN ResNet101 (Depth) & 0.0158 & 0.925 & 1.7 & 1.8 \\
        \midrule
        EgoHands: SOLOv2 ResNet50 (RGB) & 0.1025 & 0.575 & 26.4 & \textbf{29.9} \\
        EgoHands: SOLOv2 ResNet101 (RGB) & 0.0975 & 0.500 & 22.9 & 25.0 \\
        EgoHands: Mask R-CNN ResNet50 (RGB) & 0.0742 & 0.950 & 16.1 & 17.4 \\
        EgoHands: Mask R-CNN ResNet101 (RGB) & \textbf{0.1040} & 0.900 & 21.4 & 25.3 \\
        \midrule
        ObMan: SOLOv2 ResNet50 (RGB-D) & 0.0606 & 0.450 & 13.8 & 16.5 \\
        ObMan: SOLOv2 ResNet101 (RGB-D) & 0.0535 & 0.375 & 13.5 & 16.0 \\
        ObMan: Mask R-CNN ResNet50 (RGB-D) & \textbf{0.0798} & 0.975 & 18.7 & 21.7 \\
        ObMan: Mask R-CNN ResNet101 (RGB-D) & 0.0786 & 0.950 & 20.6 & \textbf{22.7} \\
        \midrule
        RHD: SOLOv2 ResNet50 (RGB-D) & 0.0591 & 0.725 & 13.0 & 14.8 \\
        RHD: SOLOv2 ResNet101 (RGB-D) & 0.0794 & 0.675 & 15.7 & 16.8 \\
        RHD: Mask R-CNN ResNet50 (RGB-D) & 0.0775 & 0.975 & 16.5 & \textbf{17.5} \\
        RHD: Mask R-CNN ResNet101 (RGB-D) & \textbf{0.0839} & 0.975 & 16.9 & 17.8 \\
        
        \bottomrule
    \end{tabular}
\end{table}

The code for adapting the dataset can be accessed on our GitHub repository. The models were trained using adapted datasets with the training parameters outlined in section 3.2. We evaluated the trained models on the test dataset, which is representative of the anticipated environment.
Many existing datasets for hand recognition rely on certain assumptions, such as the hands being the closest objects to the camera and being in the centre of the image, and the absence of other objects in the frame. In reality, these assumptions may not hold true in all situations. We also assume that they have influenced the evaluation results presented in Table \ref{tab_dataset_stats5}.

The PDQ curves generated by models trained on external datasets resemble those of our DR dataset in shape but have lower values for any given score threshold, resulting in lower maximum PDQ values. 

The DenseHands dataset showed the worst generalization, both in terms of PDQ and AP metrics. This dataset does not contain any distractors present in the scene; moreover, the images of this dataset are characterized by low variability in hand position and orientation. In contrast, the ObMan dataset, which contains several background objects, including a 3D human model, showed better results in terms of both PDQ and AP metrics. Nevertheless, the ObMan hands are located predominantly in the centre of the image while also maintaining a similar distance for the camera. The high variability of the RHD images allows the corresponding models to obtain similar results to ObMan. Nonetheless, the absence of distractors in the RHD dataset limits the quality of the predictions. The results of the models trained on RHD and ObMan are similar, but the RHD models obtain slightly better results in terms of PDQ, whereas the ObMan models obtain slightly higher APs.

Despite having the smallest sample size among all datasets, EgoHands produced the best results for both metrics (PDQ\textsubscript{max} 0.1040, AP\textsubscript{max} 25.3). However, it is also one of the two datasets based on real camera data, so we assume that this factor is the reason for such a high result. The HandSeg dataset, which was also collected with a real depth camera, shows low variability in dataset features and performs significantly worse than EgoHands in terms of both metrics. The best results for all datasets were achieved by the Mask R-CNN models.

In general, however, the results obtained by training the models on the existing datasets are much worse than the results obtained by training the models on our DR dataset, despite the fact that this dataset is randomly generated.
Overall, these findings highlight the importance of dataset characteristics, particularly the inclusion of distractors and variability in image content, for the development of effective hand detection models.

\subsection{Comparison with existing solution}

In order to compare the performance of our trained models with the state-of-the-art MediaPipe (release 0.9.0) solution we evaluate bounding box detection for both models. Since MediaPipe predicts landmark positions for each hand and SOLOv2 models output the instance masks we envelop their predictions into axis-oriented bounding boxes. MediaPipe model’s complexity was set to 1 (notating the more accurate and thus more complex model), the static image mode was enabled since the images in the dataset do not represent a video sequence and the use of MediaPipe’s tracking would not be beneficial. The maximum number of recognized hand instances was set to 2, since in our dataset the maximum number of instances per sample was also limited to 2. The AP and PDQ evaluations were performed for confidence thresholds in the range [0.0,1.0] with 0.025 steps. Evaluation results are shown in Table \ref{tab_dataset_stats6}.

\begin{table}[h!]
    \caption{AP and PDQ evaluated for bounding boxes for Mask R-CNN ResNet50 models trained on our DR dataset compared with MediaPipe. Evaluated on test dataset (real camera images)}\label{tab_dataset_stats6}
	\centering
    \begin{tabular}{lcccc}
    \toprule
    Model &
    PDQmax  &
    \begin{tabular}{p{40mm}}
    Confidence score threshold at PDQ\textsubscript{max}
    \end{tabular}
    & AP at PDQ\textsubscript{max} & AP\textsubscript{max} \\
    \midrule
    Mask R-CNN ResNet101 (Depth) & 0.1318 & 0.825 & 33.9 & 36.4 \\
    SOLOv2 ResNet50 (RGB) & 0.1147 & 0.600 & 38.0 & \textbf{47.4} \\
    Mask R-CNN ResNet50 (RGB-D) & \textbf{0.1576} & 0.900 & 42.4 & 45.5 \\
    MediaPipe & 0.0836 & 0.050 & 18.1 & 18.1 \\
    \bottomrule
    \end{tabular}
\end{table}

\begin{figure}[h!]
	\centering
\begin{subfigure}[t]{\textwidth}
\centering
    \includegraphics[width=\textwidth]{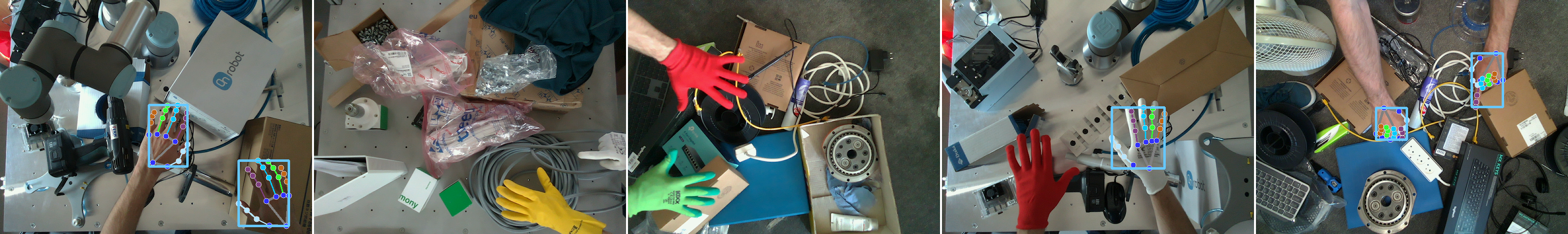}

    \label{fig_mediapipedet}
\end{subfigure}

\begin{subfigure}[t]{\textwidth}
\centering
    \includegraphics[width=\textwidth]{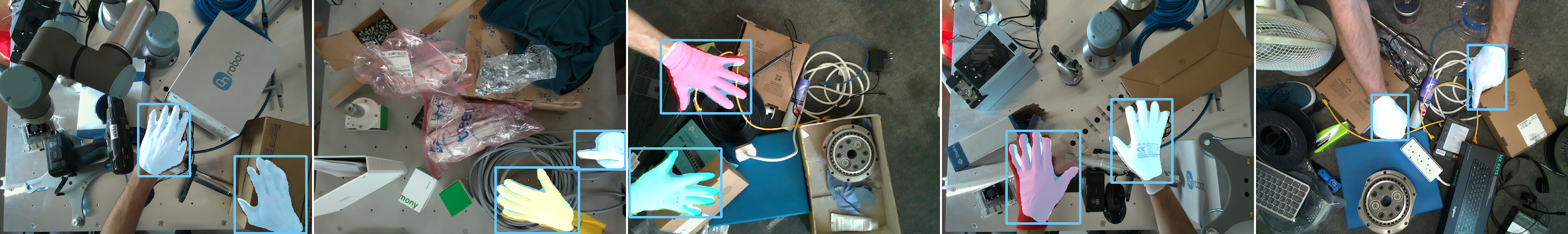}
    
    \label{fig_ourdet}
\end{subfigure}
    \caption{Side-by-side comparison of predictions: (first row) MediaPipe Hands - landmarks and bounding boxes; (second row) Mask R-CNN ResNet50 RGB-D trained on our DR dataset – masks and bounding boxes.}
    \label{fig_detections}
\end{figure}

The results of this evaluation show that a deep learning model for hand detection based on Mask R-CNN trained on a custom synthetic dataset outperforms the state-of-the-art solution, MediaPipe, in terms of both PDQ and AP metrics when evaluated on bounding boxes. Specifically, the MediaPipe solution achieved PDQ of 0.0836 and AP of 18.1, while the proposed Mask R-CNN model trained on the synthetic RGB-D dataset achieved PDQ of 0.1546 and AP of 45.5. To further understand the reason for this outcome, we made a qualitative comparison of model predictions. Examples of side-by-side predictions are available in Figure \ref{fig_detections}. 

It can be observed that MediaPipe's predictions tend to fail if work gloves are used. In addition, it is worth noting that red and yellow gloves were predicted with a higher probability than green ones, which we assume is due to the similarity of the typical skin colour to the red spectrum in the colour space. As opposed to this, our best performing model, namely Mask R-CNN ResNet50 RGB-D, predicts the instances more stably regardless of illumination conditions and work glove colour. We do not assert that our trained models are superior to Mediapipe and can fully replace it, especially considering the broad range of capabilities that Mediapipe provides. However, within our specific industrial setting, our models effectively recognize hand instances and fulfil this particular task.

\section{Conclusion}
An essential factor for achieving optimal performance in deep learning-based applications is the availability of large, accurately labelled training datasets. However, manual data collection and annotation can be a laborious and costly process. This study follows up on a topic of a synthetic dataset generation \cite{vysocky_generating_2022}, which incorporates Domain Randomization to create large and accurately annotated datasets for multimodal hand instance segmentation. The generator randomizes several scene features, including the hand pose and texture, scene texture and lighting, and simulated noise and occluding objects. Although the resulting images are crude and not photorealistic, these characteristics can encourage the deep neural network to focus on the fundamental problem structure instead of details that may be absent in real-world scenarios during testing, improving the network's ability to generalize. High data variety in samples reduces the likelihood of models overfitting to synthetically generated training datasets. Furthermore, these datasets can be created more quickly and with less expertise required.

The effectiveness of the synthetic datasets was assessed by training state-of-the-art instance segmentation models (SOLOv2 and Mask R-CNN) and evaluating their performance on a complex test dataset including real images of hands. In general, Mask R-CNN models generally outperform other models for all modalities, with the RGB-D models performing comparably with the RGB models in terms of average precision. All models trained on the DR dataset showed a significant increase in metrics compared to models trained using the COCO dataset. The SOLO models had significantly lower prediction confidence scores compared to the corresponding Mask R-CNN models, partly due to the way the confidence scores are calculated in these models, but also potentially due to differences in the characteristics of the generated DR dataset and the actual camera images. Since the choice of confidence score threshold plays a critical role in evaluating the performance of a model, the probability-based detection quality (PDQ) metric was used to search for the optimal confidence score threshold. The Mask R-CNN ResNet50 RGB-D model was found to be the best-performing model, achieving high PDQ at a high confidence score threshold while retaining high average precision. The qualitative evaluation also supports the conclusion that the RGB-D models outperform both depth and RGB models in terms of the quality of instance detection.
By comparing our DR dataset to several publicly available hand datasets, we found that many existing datasets for hand recognition rely on certain assumptions, such as the hands being the closest objects to the camera and being located in the centre of the image, and the absence of other objects in the image. The results of AP and PDQ evaluations obtained by training the models on the existing datasets (even those based on real camera data) are significantly worse than the results obtained by training the models on our DR dataset. This again highlights the importance of dataset characteristics in developing accurate hand detection models. In particular, incorporating variability in image content and the presence of random distractors can significantly improve the performance of such models, making them more suitable for real-world applications in various environments.

The trained instance segmentation models are intended to be part of the human-robot interaction interface. For this particular scenario, it is essential that the trained models recognize instances regardless of their colour, as workers often use work gloves in industrial environments. In order to compare the performance of our trained models with the state-of-the-art MediaPipe solution we evaluate both models in terms of AP and PDQ metrics for bounding boxes. The results of this evaluation show that a deep learning model for hand detection based on Mask R-CNN trained on a custom synthetic dataset outperforms the state-of-the-art solution, MediaPipe, in terms of both PDQ and AP metrics when evaluated on bounding boxes. Our qualitative comparison of model predictions showed that MediaPipe's predictions tend to fail when work gloves are used and that red and yellow gloves were predicted with higher probability than green ones. We do not assert that our trained models are superior to Mediapipe and can fully replace it, especially considering the broad range of capabilities that Mediapipe provides. However, within our specific industrial setting, our best-performing model predicts the instances more stably regardless of illumination conditions and work gloves colours. Overall, our work highlights the importance of using a high-quality, diverse dataset for training hand detection models in industrial settings and provides a promising direction for future research in this area.

The findings of this study indicate that deep convolutional neural networks can be trained to accurately segment hand instances in real-world images by utilizing solely synthetic datasets. This method has the advantage of not requiring manual annotation, a labour-intensive and time-consuming process, and of producing pixel-level accurate masks for each instance. Conversely, annotated datasets created manually are vulnerable to errors that may arise from human oversight, leading to the production of noisy data as multiple annotators may have discrepancies in their annotations. The utilization of synthetic datasets eliminates the need for collecting and annotating large training datasets for a specific object. Instead, it becomes possible to create extensive datasets for various objects of interest by utilizing their CAD models. This is particularly advantageous in specialized tasks such as bin picking and logistics, where data may be limited. It is crucial to comprehend the essential characteristics that are required to be present in the training dataset when utilizing synthetic data for deep learning. This is necessary to enable the model to generalize effectively to real-world images. For hand instance segmentation, the model must be capable of learning the features associated with the shape of the hand. Consequently, the synthetic data generator was specifically designed to incorporate both realistic and unrealistic textures to encourage the network to learn shape-related features. This helps to ensure that the model perceives real-world images as mere variations of the synthetic data on which it was trained.

It should be highlighted that the synthetic data generator employed in this study has certain limitations, such as a restricted range of hand meshes and an absence of simulation models for hand interactions with tools. Increasing the variety of hand meshes and enabling them to interact with each other could create more variations in the synthetic dataset and improve the generalization of the models trained on it. Furthermore, simulating hand interactions with a diverse set of tools may help to narrow the discrepancy between synthetic data and real-world images, and therefore present promising avenues for future research.

\section*{Declaration of Competing Interest}
The authors have no conflicts of interest to declare.

\section*{Acknowledgements}
The Research Platform for Industry 4.0 and Robotics in Ostrava Agglomeration project (project number CZ.02.1.01\slash0.0\slash0.0\slash17\_049\slash0008425) from the Operational Program Research, Development, and Education, supported this work. Additionally, the article received financial assistance from the state budget of the Czech Republic as part of the SP2023/060 research project.

\bibliographystyle{unsrtnat}
\bibliography{references}  






\end{document}